\documentclass[11pt,onecolumn,twoside] {IEEEtran}
\usepackage{graphicx}
\usepackage{epsfig}
\usepackage{amsmath,amssymb,amsfonts}
\usepackage{amsbsy}
\usepackage{graphicx}
\usepackage{graphics}
\usepackage{algorithm}
\usepackage[noend]{algorithmic}
\usepackage{subfigure}
\usepackage{cite}
\usepackage{slashbox}
\usepackage{multirow}
\usepackage{pdfsync}
\usepackage{epstopdf}
\usepackage{color}
\usepackage{amsthm}
\usepackage{boxedminipage}

\newtheorem{proposition}{{Proposition}}

\newtheorem{definition}{{Definition}}

\ifCLASSINFOpdf
 
\else
  
\fi

\begin{document}
%
\title{Distributed average consensus with quantization refinement}
\author{Dorina~Thanou, 
        Effrosyni~Kokiopoulou, 
        Ye Pu and Pascal~Frossard  
\thanks{D. Thanou, Y. Pu and P. Frossard are with Ecole Polytechnique F\'ed\'erale de Lausanne (EPFL), Signal Processing Laboratory-LTS4, CH-1015, Lausanne, Switzerland ({e-mail:\{dorina.thanou, y.pu, pascal.frossard\}@epfl.ch}).}
\thanks{E. Kokiopoulou is with Google Research, Zurich, Switzerland (e-mail: kokiopou@google.com).}}
\maketitle
\begin{abstract}
We consider the problem of distributed average consensus in a sensor network where sensors exchange quantized information with their neighbors.  We propose a novel quantization scheme that exploits the increasing correlation between the values exchanged by the sensors throughout the iterations of the consensus algorithm.  A low complexity, uniform quantizer is implemented in each sensor, and refined quantization is achieved by progressively reducing the quantization intervals during the convergence of the consensus algorithm. We propose a recurrence relation for computing the quantization parameters that depend on the network topology and the communication rate. We further show that the recurrence relation can lead to a simple exponential model for  the size of the quantization step size over the iterations, whose parameters can be computed a priori.  
Finally,  simulation results demonstrate the effectiveness of the progressive quantization scheme that leads to the consensus solution even at low communication rate. 

\end{abstract}
\begin{IEEEkeywords}
Distributed average consensus, sensor networks, progressive quantization.
\end{IEEEkeywords}

\IEEEpeerreviewmaketitle

\section{Introduction}
\label{sec:introduction}
Distributed consensus algorithms \cite{Bertsekas89} have attracted a lot of research interest due to their applications in wireless network systems. They are mainly used in ad-hoc sensor networks in order to compute the global average of sensor data in a distributed fashion, using only local inter-sensor communication. Some of their most important applications include distributed coordination and synchronization in multi-agent systems \cite{Blondel05}, distributed estimation \cite{Schizas2008}, distributed classification \cite{Kokiopoulou2011} and distributed control problems.  


While in theory convergence to the global average is mostly dependent on the sensor network topology, the performance of distributed average consensus algorithms  in practical systems is largely connected to the power or communication constraints and limited precision operations. In general, the information exchanged by the network nodes has to be quantized    
prior to transmission due to limited communication bandwidth and limited computational power. However, this quantization process induces some quantization noise that is accumulated throughout the iterative consensus algorithm and affects its convergence, leading to significant performance degradation \cite{Xiao07}.

In this paper, we design a novel distributed progressive quantization algorithm that limits the quantization noise and leads to convergence to the average value even at low bit rates. Motivated by the observation that the correlation between the values communicated by the nodes increases along the consensus iterations, we build on our previous work \cite{Thanou_progquantICASSP} and  propose to progressively and consistently reduce the range of the quantizer in order to refine the information exchanged in the network and guide the sensors to converge to the average consensus value.  The proposed quantization scheme is computationally simple and consistent throughout the iterations as every node implements the same quantization all the time.   We describe a method for computing offline the parameters of the quantizers, which depend on the network topology and the communication constraints. 
Our design is based on an average case analysis, which leads to effective performance in realistic settings. Convergence of the consensus iterations is achieved when the energy of the quantization error decreases with the iterations. 
We illustrate the performance of the proposed scheme through simulations that demonstrate that the consensus algorithm converges fast to the average value even in the case where the information is hardly quantized.

A few works have been proposed recently to address the problem of quantization in consensus averaging algorithms.  In particular, it was shown in \cite{Xiao07} that if the quantization noise is modeled as white and additive with fixed variance then consensus cannot be achieved. The authors in \cite{Aysal08} propose a probabilistic quantization scheme that is shown to reach a consensus almost surely to a random variable whose expected value is  equal to the desired average. Unfortunately, the scheme performs poorly at low bit rate. Kashyap et al. \cite{Kashyap07} designed an average consensus algorithm with the additional constraint that the states of the agents are integers.  Thus convergence is achieved to some integer approximation of the average of the initial states. Modifications of the classical consensus algorithm have been proposed in \cite{carli2009,Fang2010}, including a compensation error term that guarantees that the average is preserved at each iteration.  A coding scheme based on predictive coding is proposed in \cite{Yildiz08} in order to exploit the temporal correlation among successive iterations. Convergence to the true average is shown to be possible under the condition that the quantization noise variance vanishes over the iterations. A different approach for dealing with the quantization noise and additive white noise in general is proposed in \cite{Kar2010} and \cite{Mosquera2010} respectively. Both algorithms adapt the weight link sequence in order to guarantee convergence under certain conditions, at a cost of lower convergence rate. In general, all the above mentioned algorithms either maintain the average value in the network but cannot reach a consensus effectively, or converge to a random variable that is not always the target average value.

More recently, the authors in \cite{Fang09}, \cite{carli} and \cite{Li-2011} have  proposed quantization strategies that maintain the average of the initial state and at the same time converge asymptotically to the true average value. The quantization scheme introduced in \cite{Fang09} adaptively adjusts the quantization step-size by learning from previous states, at the price of significant complexity and memory requirements. The quantization threshold in \cite{carli} is  adapted online based on a zoom-in zoom-out strategy, while the set of quantization levels is maintained constant over the iterations. Although these last two solutions perform quite well at high bit rates, the convergence rate appears to be slow when the quantization is coarse. In addition, the stability of both quantization schemes depends on the choice of globally defined parameters that are not easy to determine a priori.  Finally, the scheme proposed in  \cite{Li-2011}  is the one that is closer to our work since it is based on the assumption that, as consensus is achieved asymptotically the prediction error of the sensors' states tends to zero. The scaling function is selected in such a way that it decays over time without causing the saturation of the quantizer. However, the proposed scheme leads to very conservative selection of the parameters of the scaling function that prevents reading significant gains in the average consensus performance. On the other hand, our system is able to achieve fast convergence to the true average value in realistic settings, even if it does not provide strict performance guarantees; it relies on average case analysis instead of conservative bounds.  

The structure of the paper is as follows. Section \ref{ProgAlg} presents our new progressive quantization algorithm. A recursive method for computing the quantization parameters is proposed in Section \ref{ParameterDesign}, followed by a simple exponential relation for adapting the quantizer step-size. Simulations results are finally presented in Section \ref{simulations}.

\section{Progressive quantizer for distributed average consensus}
\label{ProgAlg}

We consider a sensor network topology that is modeled as a weighted, undirected graph $\mathcal{G}=(V,\mathcal{E})$, where $V \in \{ 1, \ldots, m \} $ represents the sensor nodes and $m=|V|$ denotes the number of nodes. An edge denoted by an unordered pair $\{i,j \}\in{\mathcal{E}}$, represents a link between two sensor nodes $i$ and $j$ that communicate with each other. Moreover,  a positive weight $W(i,j)>0$ is assigned to each edge if $\{i,j \}\in{\mathcal{E}}$. The set of neighbors for node $i$ is finally denoted as $\mathcal{N}_i=\{j|\{i,j\}\in\mathcal{E}\}$. 

The node states over the network at time $t$ can be expressed as a vector $z_t=[z_t(1), \ldots, z_t(m)]^T$, where $z_t(i)$ represents a real scalar assigned to node $i$ at time $t$. The distributed average consensus problem consists in computing iteratively at every node the average $\mu =\frac{1}{m}\sum_{i=0}^m z_0(i)$, where $z_0(i)$ is the initial state at sensor $i$. The consensus can be achieved by linear iterations of the form $z_{t+1}= W z_t$, where the symmetric weight matrix $W$  satisfies the conditions that are required to achieve
asymptotic average consensus \cite{Xiao04}, expressed as 
\begin{equation} \label{eq:condW}
\boldsymbol{1}^TW=\boldsymbol{1}^T,\ W\boldsymbol{1}=\boldsymbol{1},\ \rho{(W-\boldsymbol{1}\boldsymbol{1}^T/m)}<{1},
\end{equation}
with $\rho(\cdot)$ the spectral radius of the matrix and $\boldsymbol{1}$ is the vector of ones. 

When communication rate is limited between sensors, the value ${z}_{t}(i)$ of a sensor node $i$ at each step $t$ is quantized prior to its transmission to neighbor nodes. The quantized value $\hat{z}_{t}(i)$ can be written as 
\begin{equation}\label{eq:quantZ}
\hat{z}_{t}(i)={z}_{t}(i)+\epsilon_t(i),
\end{equation}
where $\epsilon_t(i)$ models the additive quantization noise of sensor $i$ at iteration $t$. In particular, in the case of a $n$-bit uniform quantizer, the quantized values can be written as $ \hat{z}_t(i)=\left \lfloor \frac{z_t(i)-z^{(\min)}_0}{\Delta} \right\rfloor \cdot\Delta+\frac{\Delta}{2}+z^{(\min)}_0,$ when the initial sensor states lie in a finite interval  of size $S=z^{(\max)}_0-z^{(\min)}_0$. The parameters  $z^{(\min)}_0$ and  $z^{(\max)}_0$ represent the minimum and the maximum values of the interval respectively. The parameter $\Delta=S/2^{n} $ is the quantization step-size, which drives the error of the quantizer.

In the presence of quantization noise in the distributed average consensus algorithm,  we use the following linear iterations that preserve the  average of the initial states \cite{carli2009}
\begin{equation}
z_{t+1}=z_t+(W-I)\hat{z}_t,
\label{eq5}
\end{equation}
where $I$ is the identity matrix. An analytical expression of Eq.(\ref{eq5}) shows that the quantization error propagates through the iterations of the consensus algorithm. More specifically, the states $z_{t+1}$ and  $\hat{z}_t$ are expressed as
\begin{equation}
{\small
\hat{z}_{t}=W^{t}z_0+\sum_{s=0}^{t-1}W^s(W-I)\epsilon_{t-s-1}+\epsilon_t 
\label{eq11}
}
\end{equation}

\begin{equation}
{\small
z_{t+1}=W^{t+1}z_0+\sum_{s=0}^{t}W^s(W-I)\epsilon_{t-s}.
\label{eq12}
}
\end{equation}

The linear iterations defined by Eq.(\ref{eq5})  preserve the average of the initial states in the network but unfortunately do not guarantee convergence to the average consensus value in the presence of quantization.  In order to limit the influence of the quantization noise, we should decrease the step size by either increasing the number of bits or  adapting the quantization range for the same number of bits. In the average consensus problem, it can be observed that as the number of iterations increases, the correlation between the sensors' states increases and the values computed by the sensors tend to converge into an interval of decreasing size. Quantization in the full range of size $S$ hence results in a waste of bits or in limited precision that prevents the algorithm to converge to the true average value. We therefore propose to adapt the quantization step-size as the number of linear iterations increases in a new progressive quantization algorithm. We keep a simple uniform quantizer with a fixed number of bits per sensor and we adapt the quantization range so that quantization becomes finer along the iterations. 

In more details, we denote the size of the range of the quantizer in node $i$ at time $t$ as $S_t(i)$. Since the size of the quantization range is always positive, we impose $S_t(i)>0$. This range decreases in each sensor as the iterations proceed. The quantization range is further centered around the previous state of the consensus algorithm $\hat{z}_{t-1}(i)$ as the values of the consensus algorithm converge over time. More formally, the sensor $i$ encodes its state $z_{t+1}(i)$ by using a quantization interval that is defined as $[\hat{z}_{t}(i)-S_{t+1}(i)/2,  \hat{z}_{t}(i)+S_{t+1}(i)/2]$. The data is uniformly quantized in this reduced range, which leads to a step-size 
\begin{equation}\label{eq:Delta_t}
\Delta_{t+1}=\frac{S_{t+1}(i)}{2^{n}}
\end{equation} 
that decreases over time. The values falling out of the quantization interval are clipped and coded to the nearest quantizer value. In order to simplify the design of the quantizer for realistic settings, we impose the size of this interval to be identical for all the sensors, independently of their previous state and their position in the network (i.e., $S_t(i) = S_t, \ \ \forall i=1,\ldots,m$). Since each neighbor node $j\in \mathcal{N}_i$ knows the value $\hat{z}_{t}(i)$ received at the previous iteration, it is able to perform inverse quantization and to compute correctly the value $\hat{z}_{t+1}(i)$. We call the proposed quantization scheme Progressive Quantizer. 

The important parameter in the Progressive Quantizer algorithm is clearly the size $S_t$ of the quantization range. It has to be small enough such that the precision of the quantized information is sufficient  for convergence to the true average value. On the other hand, it should be chosen large enough such that the values computed in the network nodes fall in the quantization range with high probability in order to avoid clipping that may negatively affect the convergence of the average consensus algorithm. 

\section{Design of the parameters of the progressive quantizer}
\label{ParameterDesign}
\subsection{Average case analysis}

In this section, we propose a constructive methodology to compute a priori the size of the quantization range, based on the properties of the network topology and the communication constraints. 
In order to guarantee that the quantizer does not saturate (i.e., all the values fall inside the quantization range), $S_{t+1}$ should satisfy the following inequality
\begin{equation}
{\small
 \|z_{t+1}-\hat{z}_{t}\|_\infty\le \frac{S_{t+1}}{2}.
 \label{infineq}
}
\end{equation}
 The computation of the worst case interval based on (\ref{infineq}) typically leads to  conservative progressive quantizer design that does not necessarily lead to to a better performance than the classical uniform quantizer with a constant range. This observations is supported by simulations in subsection \ref{otherschemes}  where a conservative design  \cite{Li-2011} is unable to lead to fast convergence.  Instead of looking for strong guarantees such as those in  (\ref{infineq}), we build our quantizer such that values fall in the quantization range with high probability. This comes at a price of some potential clipping, which however does not significantly penalize the convergence of the algorithm.  Moreover, limiting a priori the dynamic range of the sensors' states in a meaningful way is expected to prevent the consensus algorithm from being affected by potential outliers that  occur due to  quantization noise. The overall procedure could be characterized as an attempt to guide the quantized consensus algorithm such that it converges to the average value  with a rate that gets close to the convergence rate of the ideal unquantized consensus algorithm.  
  
In more details,  we propose to relate the quantizer range size to  the mean square difference 
$\|z_{t+1}-\hat{z}_t\|^2/m$  between two successive values. It leads to the following average case condition 
\begin{equation}
{\small
\frac{E[\|z_{t+1}-\hat{z}_t\|^2] }{m}\le \big(\frac{S_{t+1}}{2}\big)^2,
\label{eq10}
}
\end{equation}

where $\| \cdot \|$ denotes the L2 norm.   The expectation of the difference between successive values in the consensus algorithm represents the minimal size of the quantization range at each iteration. Moreover, we model the quantization noise samples as spatially and temporally independent random variables that are uniformly distributed with zero mean and variance $\Delta_{t}^2/12$. 
In the sequel, we  first derive in Proposition \ref{Prop1} an upper-bound of $E[\|z_{t+1}-\hat{z}_t\|^2]$ that depends on the  previous values $\{S_1,...,S_{t}\}$. This upper-bound together with (\ref{eq10}) permits to estimate $S_{t+1}$. 
\begin{proposition}\label{Prop1} Let $\hat{z}_t$ and $z_{t+1}$ be defined as in Eqs. (\ref{eq11}), (\ref{eq12}). Let also $\lambda_2$ be defined as $\lambda_2:=\rho(W-\frac{{\mathbf1\mathbf{1}^T}}{m})$ and $\lambda_{\min}$ be the smallest algebraically eigenvalue of $W$. Then, 
it holds that
\begin{equation}
{\small
\begin{split}
E[\|z_{t+1}-\hat{z}_{t} \|^2]& \le \|z_0\|^2 \lambda_2^{2t} (1-\lambda_{\min})^2 +(1-\lambda_{\min})^2 \sum_{s=0}^{t-1}\|W^s(W-I)\|^2m\frac{S_{t-s-1}^2}{2^{2n}\cdot 12}
\\&+ (2-\lambda_{\min})^2m\frac{S_{t}^2}{2^{2n}\cdot 12}.
\end{split}
\label{eq:Prop}
}
\end{equation}
\end{proposition}
The proof of Proposition \ref{Prop1} is given in Appendix \ref{app1}. 

Then, Eqs. (\ref{eq10}) and (\ref{eq:Prop}) along with the fact that $\|z_0\|^2 \le m \|z_0\|_{\infty}^2$, imply that
{\small
\begin{equation}
\begin{split}
( \frac{S_{t+1}}{2})^2&=\|z_0\|_{\infty}^2\lambda_2^{2t} (1-\lambda_{\min})^2 +(1-\lambda_{\min})^2 \sum_{s=0}^{t-1}\|W^s(W-I)\|^2\frac{S_{t-s-1}^2}{2^{2n}\cdot 12}
\\ &+ (2-\lambda_{\min})^2 \frac{S_{t}^2}{2^{2n}\cdot 12}, \quad t \ge 1.
\end{split}
\label{bound}
\end{equation}
}
The computation of the quantizer range size with (\ref{bound}) implies a recursive computation of $S_t$ at each time step $t$ of the consensus algorithm. We first set $S_0$ according to the initial range of the quantizer i.e., $S_0=z^{(\max)}_0-z^{(\min)}_0$ and we compute $S_1$ from a simplified version of (\ref{bound}) where the intermediate term from the right hand side is dropped. Then  $S_{t+1}$ is computed recursively according to  (\ref{bound}), where only positive solutions are kept. 

Finally, we note that the terms used in the recursive computation of the quantization range reflect the characteristics of the network and the communication constraints. 
 The values of the estimated quantization range depend on the convergence rate of the average consensus algorithm in the absence of quantization noise $\lambda_2$, on the maximum value of the initial data $\|z_0\|_{\infty}$,  on the network topology $W$ (through $\lambda_2, \lambda_{\min}$) and on the number of quantization bits  $n$ for each sensor. Moreover, we exploit the properties of the weight matrix $W$ by taking into account the averaging effect over the successive iterations. We show in the next section that the recursive computation of the quantization range size can be approximated with a  simple exponential model.

\subsection{Exponential quantization range}\label{sec:linearmodel}

We build on the convergence behavior of the consensus algorithm and propose an approximate exponential model for the computation of the size of the quantization range. 
 We first pose without loss of generality that $S_t =2\cdot e^{-\beta_{t}}$.  
 In what follows, we show that under a few simplifying assumptions the recursive relation of the previous section leads  to an exponential model whose parameters can be determined in closed form. This closed form parameter computation is of great benefit towards deployment in realistic settings. 
 
When $S_t =2\cdot e^{-\beta_{t}}$, Eq.(\ref{bound}) becomes 
{\small
\begin{equation}
\begin{split}
e^{-2\beta_{t+1}}&=\|z_0\|_{\infty}^2\lambda_2^{2t} (1-\lambda_{\min})^2 +(1-\lambda_{\min})^2 \sum_{s=0}^{t-1}\|W^s(W-I)\|^2\frac{e^{-2\beta_{t-s-1}}}{2^{2n}\cdot 3}
\\ &+ (2-\lambda_{\min})^2 \frac{e^{-2\beta_{t}}}{2^{2n}\cdot 3}, \quad t \ge 1.
\end{split}
\label{boundexp1}
\end{equation}
}
The second term in the right-hand side of Eq.(\ref{boundexp1}) is due to the accumulated quantization error from the previous $t$ iterations. 
In particular, the matrix norm $\|W^t(W-I)\|$ that multiplies the quantization noise vectors decays asymptotically to zero. In addition to that, the sequence defined by Eq.(\ref{boundexp1})  decays to zero as specified by the following proposition. 

\begin{proposition}\label{Prop3} Let  $e^{-\beta_{t+1}}$ be a sequence defined by Eq.(\ref{boundexp1}). If the condition $\frac{(1-\lambda_{\min})^4}{(1-\lambda^{2}_2)} +(2-\lambda_{\min})^{2} < 3\cdot 2^{2n}$ is satisfied,  then $\lim_{t\rightarrow \infty} e^{-\beta_{t+1}}=0$.
\end{proposition}
The Proposition \ref{Prop3}, whose proof is given in  Appendix \ref{app2}, relates the decay and convergence of the size of the quantizer range to the characteristics of the network (through $\lambda_2, \lambda_{\min}$) and to the number of bits used in the quantization. Eventually, it means that the expected squared difference between consecutive values in the consensus algorithm as defined by the recursive equation (\ref{boundexp1}) goes to zero as long as the number of bits satisfies $n>\frac{\log{[ \frac{1}{3} (\frac{(1-\lambda_{\min})^4}{(1-\lambda^{2}_2)} +(2-\lambda_{\min})^{2}) ] }}{2\log{2}}$. Then, the following proposition shows that the second term in the right-hand side of Eq.(\ref{boundexp1}) becomes negligible  when the number of iterations increases. 


\begin{proposition}\label{Prop4} Let $W$ be a matrix satisfying the conditions defined in (\ref {eq:condW}). Let $e^{-\beta_0}, e^{-\beta_1},...,e^{-\beta_{t}}$  be a sequence such that $e^{-\beta_t}\le \delta, \forall t$ and  $\lim_{t\rightarrow \infty} e^{-\beta_{t}}=0$. Then  $\sum_{s=0}^{t-1}\|W^s(W-I)\|^2{e^{-2\beta_{t-s-1}}}$ converges asymptotically to zero for $t\to\infty$. 
\end{proposition}
The proof or Proposition \ref{Prop4} is given in Appendix \ref{app3} and it is based on the proof of Lemma 5.1 in \cite{Tetikonda2004}.  

Due to the presence of the factor $1/2^{2n}$, the second term of the right-hand side of Eq.(\ref{boundexp1}) decreases faster to zero when the quantization rate is high. This is expected, as this term captures the propagation of the quantization error. Eq. (\ref{boundexp1}) tend to follow an exponential model in this case. We can thus pose $\beta_t = \alpha \cdot t + \gamma$, which leads to an exponential decay of the size of the quantization range.  We assume that there exists an iteration $t_0$ where the second  term of the right-hand side of Eq. (\ref{boundexp1})  becomes zero. Under this assumption, Eq.(\ref{boundexp1}) simplifies to

 \begin{equation}
 {\small
 e^{-2\beta_{t+1}}=\|z_0\|_{\infty}^2\lambda_2^{2t}(1-\lambda_{\min})^2
+(2-\lambda_{\min})^2 \frac{e^{-2\beta_t}}{2^{2n}\cdot 3}, ~~ t \geq t_0,
\label{simplebound}
}
 \end{equation}
and by substituting  $\beta_t = \alpha \cdot t + \gamma$ we obtain
  \begin{equation}
 {\small
 e^{-2(\alpha \cdot (t+1) + \gamma)}=\|z_0\|_{\infty}^2\lambda_2^{2t}(1-\lambda_{\min})^2
+(2-\lambda_{\min})^2 \frac{e^{-2(\alpha \cdot t + \gamma)}}{2^{2n}\cdot 3}, ~~ t \geq t_0.
\label{simplebound2}
}
 \end{equation}
Since $\alpha$ and $\gamma$ are constant over the iterations, we choose to determine them from later iterations of the consensus (i.e., $t \geq t_0$).
First, we turn $\lambda_2$ into an exponential form by determining $\alpha$ such that $\lambda_2^{2t}=e^{-2\alpha\cdot t}$ holds. This leads to 
\begin{equation}
{\small
\alpha=-\log(\lambda_2).
}
\label{alpha}
\end{equation}
Eq. (\ref{simplebound}) then becomes
\begin{eqnarray}
 e^{-2\beta_{t+1}} & = & \|z_0\|_{\infty}^2 e^{-2\alpha\cdot t}(1-\lambda_{\min})^2 +(2-\lambda_{\min})^2\frac{e^{-2\beta_{t}}}{2^{2n}\cdot 3} \label{simplebound2} \nonumber \\
 & = &  e^{-2\beta_{t}}\left( \|z_0\|_{\infty}^2e^{2\gamma}(1-\lambda_{\min})^2 +\frac{(2-\lambda_{\min})^2}{2^{2n}\cdot 3}\right),
 ~~ t \geq t_0.
\label{simplebound3}
 \end{eqnarray}
Since $\beta_t$ is linear in $t$, we observe that 
$e^{-2\beta_{t+1}}=e^{-2(\alpha\cdot(t+1)+\gamma)}=e^{-2\alpha}e^{-2\beta_t}$,
\label{lincond}
which permits to simplify (\ref{simplebound3}) to
\begin{equation}
e^{-2\alpha}=\|z_0\|_{\infty}^2e^{2\gamma}(1-\lambda_{\min})^2
+\frac{(2-\lambda_{\min})^2}{2^{2n}\cdot 3}.
\notag
\end{equation}
We finally determine $\gamma$ as 
\begin{equation}
{\small
\gamma=\frac{1}{2}\log \left(\lambda_2^2-\frac{(2-\lambda_{\min})^2}{2^{2n}\cdot 3}\right)-\log \left(\|z_0\|_{\infty}(1-\lambda_{\min}) \right).
\label{gamma}
}
\end{equation}

We observe that the decay rate $\alpha$ of the exponential function $e^{-(\alpha\cdot t+\gamma)}$ depends on $\lambda_2$, which characterizes the convergence rate of the average consensus algorithm in the case of non-quantized communication. On the other hand, the parameter $\gamma$, apart from the eigenvalues of $W$, depends also on the number of quantization bits. It is interesting to note that the parameters of the exponential model  are similar to the parameters $\alpha', \gamma'$ that characterize the difference between two consecutive time steps in the unquantized consensus problem. In this case, the Euclidean difference between two consecutive time steps is guaranteed to be reduced by the factor  $\|W-\boldsymbol{1}\boldsymbol{1}^T/m\|<1$ at each iteration i.e.,  
\begin{equation}
{\small
\begin{split}
\frac{\|z_{t+1}-z_t\|}{\sqrt{m}}&=\frac{\|Wz_{t}-Wz_{t-1}\|}{\sqrt{m}} \le\frac{\|W-\frac{\boldsymbol{1}\boldsymbol{1}^T}{m}\|\|z_{t}-z_{t-1}\|}{\sqrt{m}}\\
&\le\frac{1}{\sqrt{m}}\|W-\frac{\boldsymbol{1}\boldsymbol{1}^T}{m}\|^t\|z_{1}-z_{0}\|\le\frac{1}{\sqrt{m}}\|W-\frac{\boldsymbol{1}\boldsymbol{1}^T}{m}\|^t\|W-I\|\|z_{0}\|\\
& \le \|W-\frac{\boldsymbol{1}\boldsymbol{1}^T}{m}\|^t\|W-I\|^2\|z_0\|_{\infty},
\label{nonquantbound}
\end{split}
}
\end{equation}
where we observe an exponential decrease over time of the form $e^{- (\alpha'\cdot t + \gamma')}$, where $\alpha'=-\mbox{log}(\|W-\frac{\boldsymbol{1}\boldsymbol{1}^T}{m}\|)$ and $\gamma'=-\log  \left(|\|z_0\|_{\infty}|\|W-I\| \right)$.
 In particular, the rate at which the exponential function decays is the same in both cases ($\alpha'=\alpha=-\log(\lambda_2)$) while, in the case of quantization, the initial value $\gamma$ of this decay depends on the number of bits (which is not the case for $\gamma'$). 
 
 In summary, with our progressive quantizer based on the exponential decay of the quantization range, we force  the average difference in the sensors'  states between two consecutive steps to decay with the same rate as in the ideal communication. At the same time,  we allow the initial magnitude of this difference to be higher for a smaller bit rate in order to take into consideration the quantization error. 
We emphasize that the above exponential model for $S_t$ yet reduces  the complexity of the proposed Progressive Quantizer. Instead of estimating a priori the values of the quantization range for a large number of iterations, the system can simply use the parameters $\alpha$ and $\gamma$ of the exponential model.  
\section{Simulation results}
\label{simulations}

In this section we analyze the performance of the Progressive Quantizer in different settings. We consider a network  of 40 sensors (i.e., $m=40$) following the random geographic graph model, i.e., the sensors are uniformly random distributed over the unit square $[0,1]\times[0,1]$.  We assume that  two neighbor sensors are connected if their Euclidean distance is less than $r=\sqrt{(\log m)/m}$, which ensures  graph connectivity with high probability \cite{GuptaKumar.00}. We consider static network topologies, which means that the edge set does not change over the iterations. As an illustration, we consider the Metropolis and the Laplacian weight matrices \cite{Xiao04} 
defined respectively as:
\begin{itemize}
\item Metropolis weights
\begin{equation}
\label{metropolis} W[i,j]= \left\lbrace
\begin{array}{ll}
		\frac{1}{1+\max\{d(i),d(j)\}}, & \mbox { if } \{i,j \}\in{\mathcal{E}}
\\ 1-\sum_{(i,k)\in\mathcal{E}}W[i,k], & \mbox { if } i=j
\\0, & \mbox { otherwise,}
\end{array}
\right.
\end{equation} 
where $d(i)$ denotes the degree of the $i^{th}$ sensor.
\item Laplacian weights
\begin{equation}
 \label{laplacian} W= I-a L
\end{equation} 
where $L$ denotes the Laplacian matrix of the graph $\mathcal{G}$ and the scalar $a$ must satisfy $0<a<1/d_{{max}}$, where $d_{max}$ consists of the maximum degree in the network.
\end{itemize}
Moreover, the initial states of the sensors are uniformly distributed in the range $[0,1]$. 


\begin{figure}[tb]
\begin{center}
\subfigure[]{\includegraphics[width=0.45\textwidth]{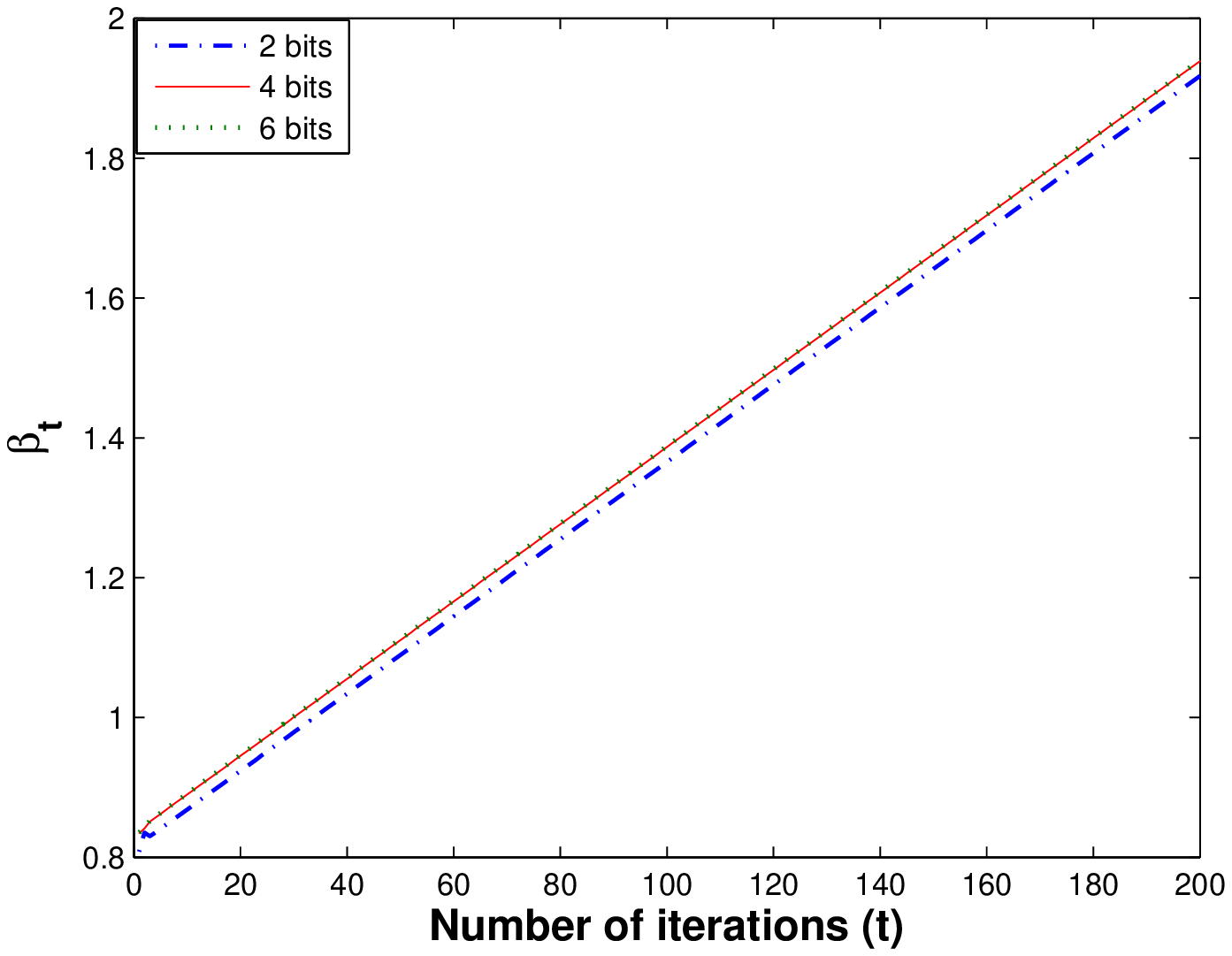}\label{fig:betavalue} }
\subfigure[]{\includegraphics[width=0.45\textwidth]{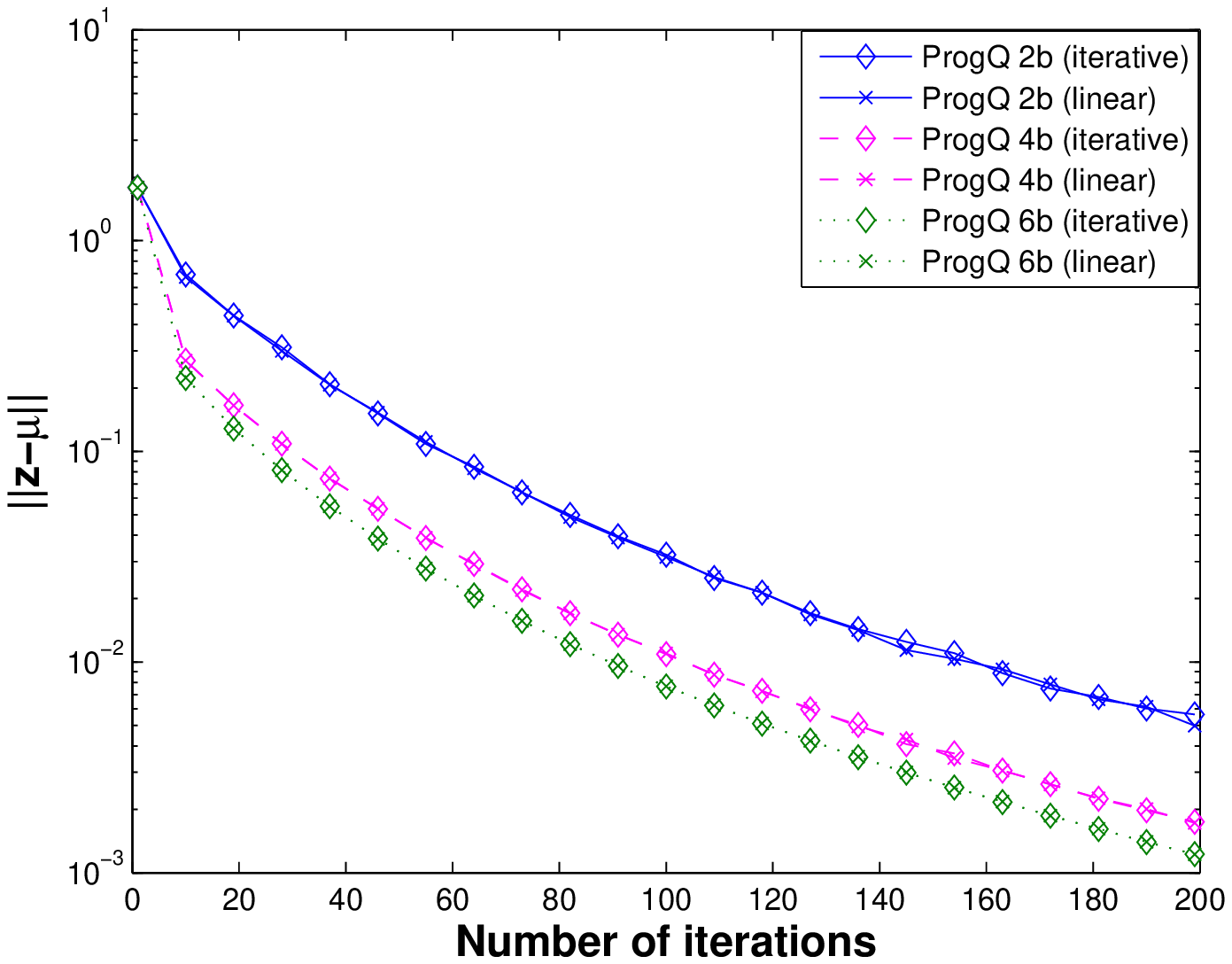}\label{fig:comparison}}
\end{center}
\vspace{-0.2cm}
\caption{(a) Evolution of the $\beta_t$ values over the iterations. (b) Comparison of the average consensus performance of the Progressive Quantizer with parameters generated (i) recursively and (ii) with a linear model.}
\vspace{-0.2cm}
\end{figure}
\subsection{Performance of the approximate exponential model}
\label{sub:compar_approx}
We first validate the exponential model for $S_t$ and we use the Metropolis weight matrix for this experiment. We compute recursively the values $S_t$ from  Eq.(\ref{bound}) for 200 random  realizations of a random network topology and communication rates of $n=[2,4,6]$ bits. For implementation issues, we fix a parameter $\delta=10^{-16}$. At iteration $t$, if the quantization range $S_t$ is smaller than $\delta$,  we quantize with the range computed at the previous iteration i.e., we set $S_t=S_{t-1}$.  All the reported experimental results are averaged over the 200 random realizations of the network topology.  In order to compare directly with the proposed exponential model in Section \ref{sec:linearmodel}, we plot the values $\beta_t$ that occur when we express the quantization range, computed  from  Eq.(\ref{bound}),  as $S_t =2\cdot e^{-\beta_{t}}$. We observe in Fig.\ref{fig:betavalue} that the value of $\beta_t$ appears to increase linearly with the number of iterations, which means that the quantization range follows an exponential function that decreases over time. Moreover, the slope of the function $\beta_t$ is independent of the bitrate, while the y-intercept value depends on the number of quantization bits. This is consistent with our approximate model (see Section \ref{sec:linearmodel}) and Eqs. (\ref{alpha}) and (\ref{gamma}), which shows that the slope $\alpha$ is dependent on the convergence rate $\lambda_2$ (and hence the network topology) and that the parameter $\gamma$ is influenced by the communication rate. We further compare the performance of the Progressive Quantizer whose parameters are computed using the approximate linear model (from Eqs. (\ref{alpha}) and (\ref{gamma})) with the performance achieved when $S_t$ is computed recursively from Eq.(\ref{bound}). Fig.(\ref{fig:comparison}) shows the obtained results. We observe that the performance is rather similar in both cases. 
This implies that the solutions of Eq.(\ref{bound}) can be well approximated with an exponential model whose parameters are easily computed. For this reason, in the rest of our experiments we adopt the approximate exponential model and compute the parameters $\alpha$ and $\gamma$ of the Progressive Quantizer using Eqs. (\ref{alpha}) and (\ref{gamma}).

\subsection{Comparison to uniform quantization}

\begin{figure*}[tb]
\begin{center}
\subfigure[Metropolis weights]{\includegraphics[width=0.45\textwidth]{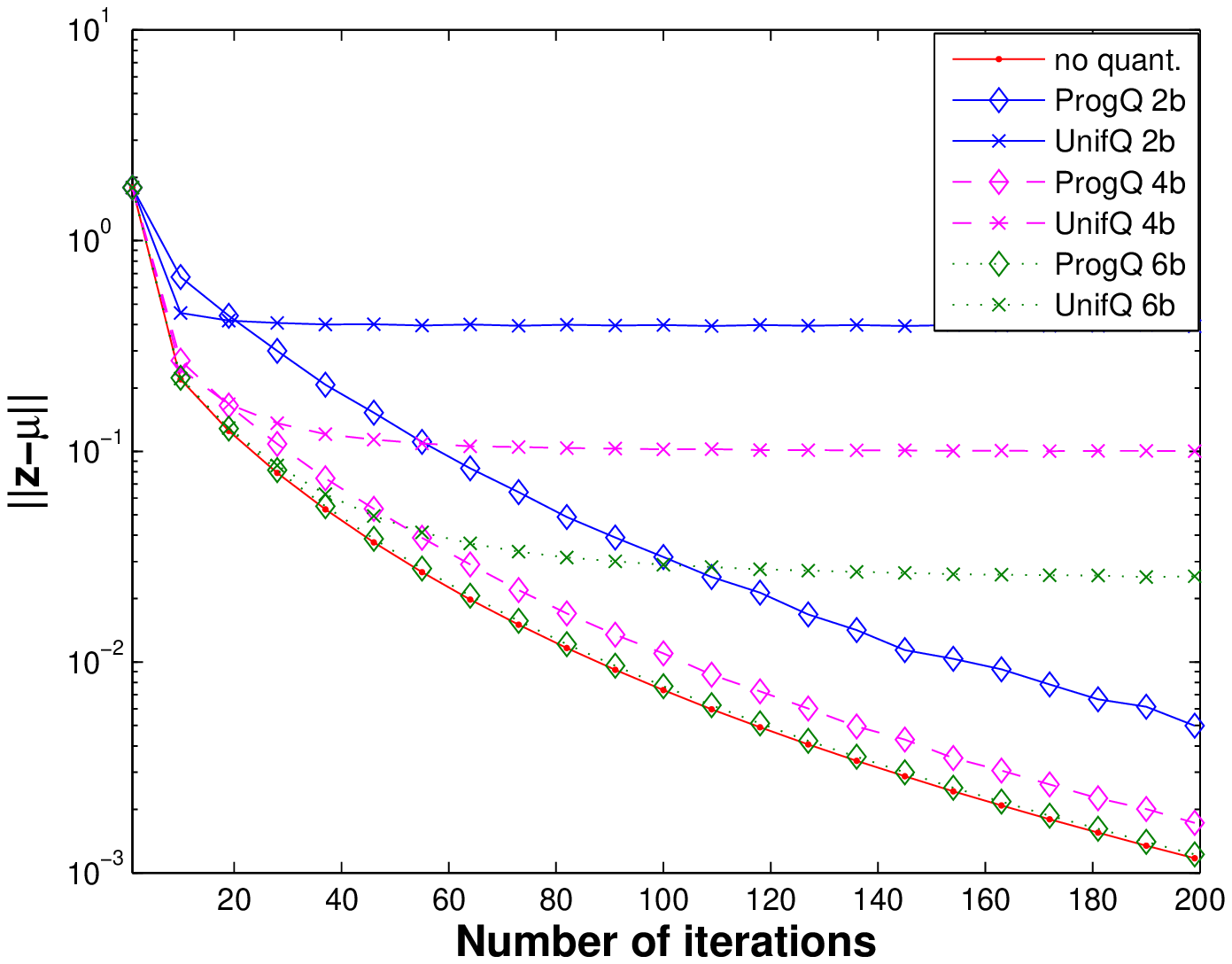}}
\subfigure[Laplacian weights]{\includegraphics[width=0.45\textwidth]{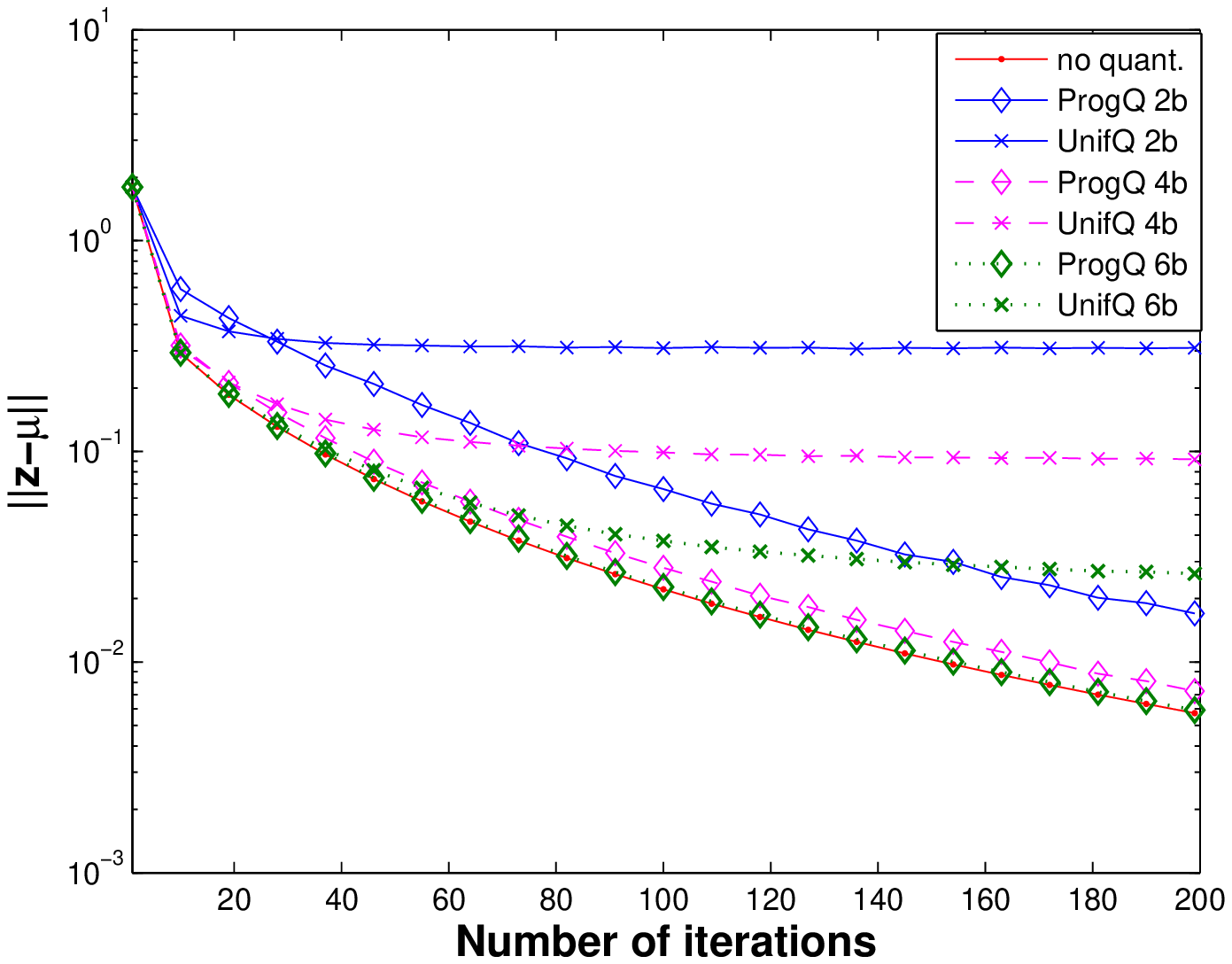}}
\end{center}
\vspace{-0.2cm}
\caption{ Average consensus performance of the proposed quantization scheme (ProgQ) vs uniform quantizer with a constant range (UnifQ) for 2, 4 and 6 bits. }
\label{fig:comp1}
\vspace{-0.2cm}
\end{figure*}

We compare the proposed quantization scheme ('ProgQ') with a baseline uniform quantization with a constant range $S=1$ ('UnifQ'), for both the Metropolis and the Laplacian weight matrices. Fig. \ref{fig:comp1} illustrates the average consensus performance corresponding to the absolute error $\|z_t-\mu \boldsymbol{1}\|_2$  versus the number of iterations for $n= [2,4,6] \mbox{ bits}$.  In order to obtain statistically meaningful results we average the error over 200 random realizations of the network topology with random  initial values. Observe that the performance of the proposed quantization scheme is very satisfactory even at a very low bit rate (2 bits). In particular, the error $\|z_t-\mu \boldsymbol{1}\|_2$ shows a decreasing behavior over the iterations, which means that the quantizer does not saturate.  It rather follows the evolution of the average consensus algorithm in the noiseless case ('no quant.'). On the other hand, the performance of the uniform quantizer with a constant range saturates quickly even at  high bit rate. 

\subsection{Comparison to existing quantization schemes for average consensus}
\label{otherschemes}

\begin{figure*}[tb]
\begin{center}
\subfigure[Metropolis weights]{\includegraphics[width=0.45\textwidth]{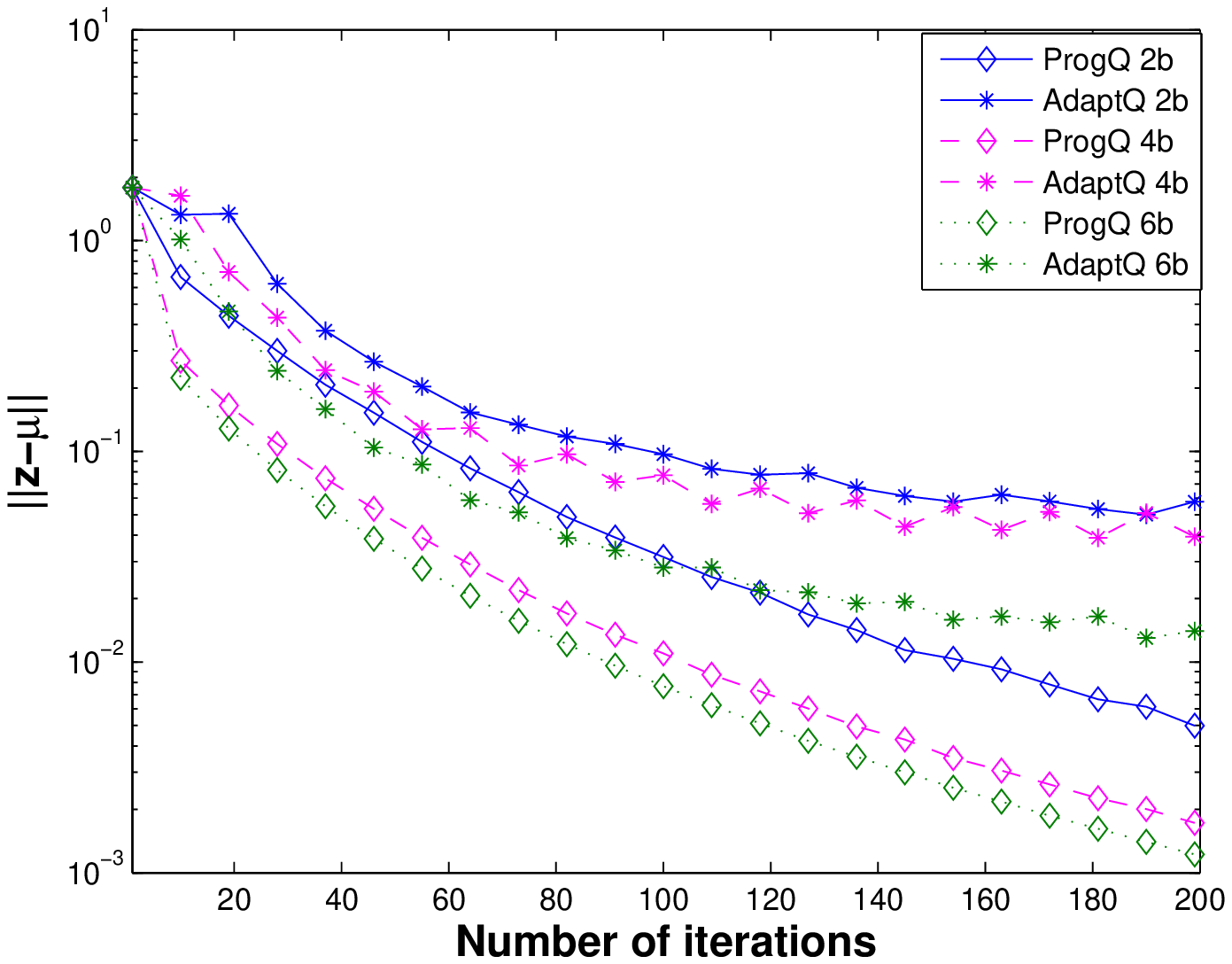}}
\subfigure[Laplacian weights]{\includegraphics[width=0.45\textwidth]{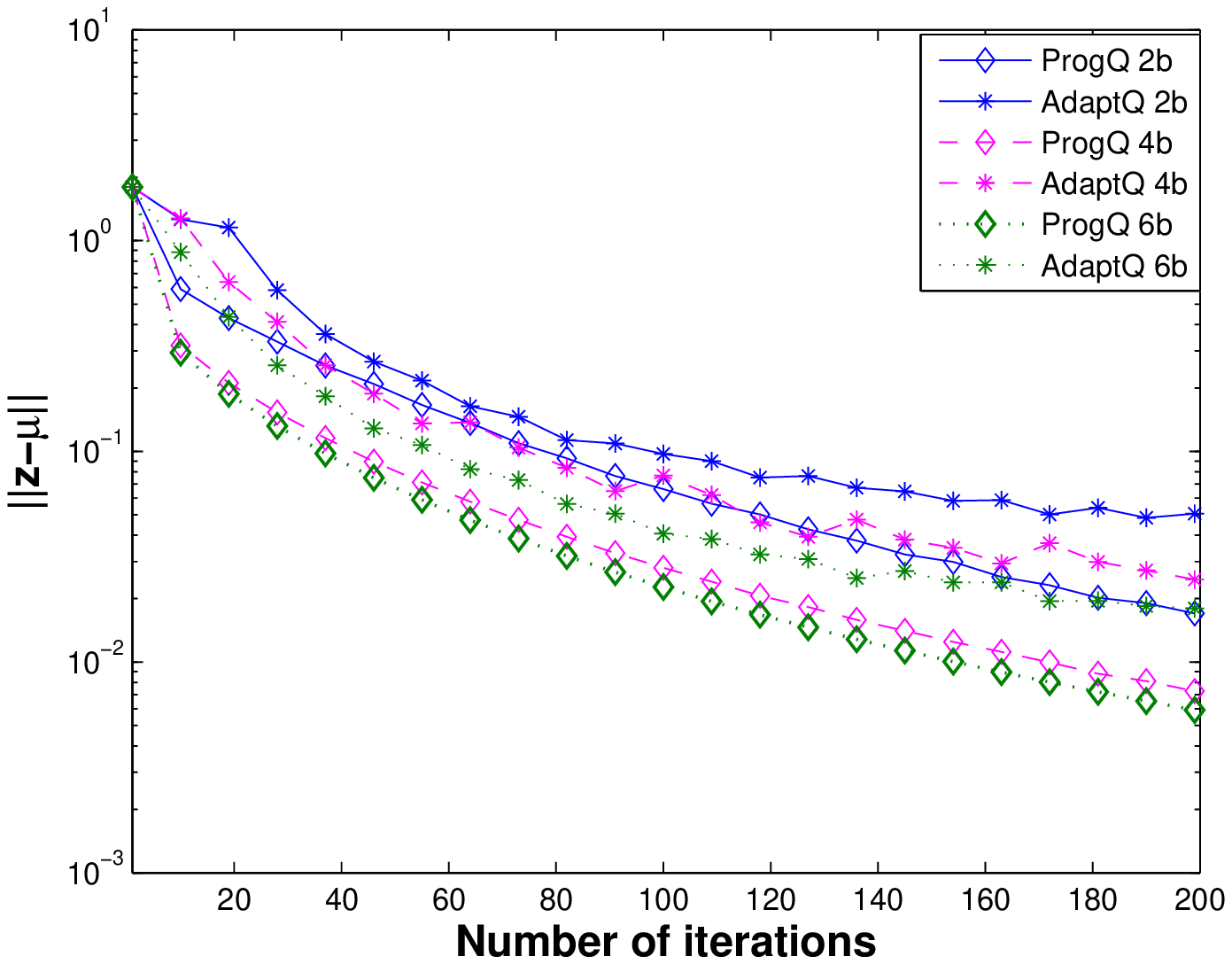}}
\end{center}
\vspace{-0.2cm}
\caption{Average consensus performance of the proposed quantization scheme (ProgQ) vs the adaptive quantizer (AdaptQ) \cite{Fang09} for 2, 4 and 6 bits.} 
\label{fig:comp2}
\vspace{-0.2cm}
\end{figure*}

\begin{figure*}[tb]
\begin{center}
\subfigure[Metropolis weights]{\includegraphics[width=0.45\textwidth]{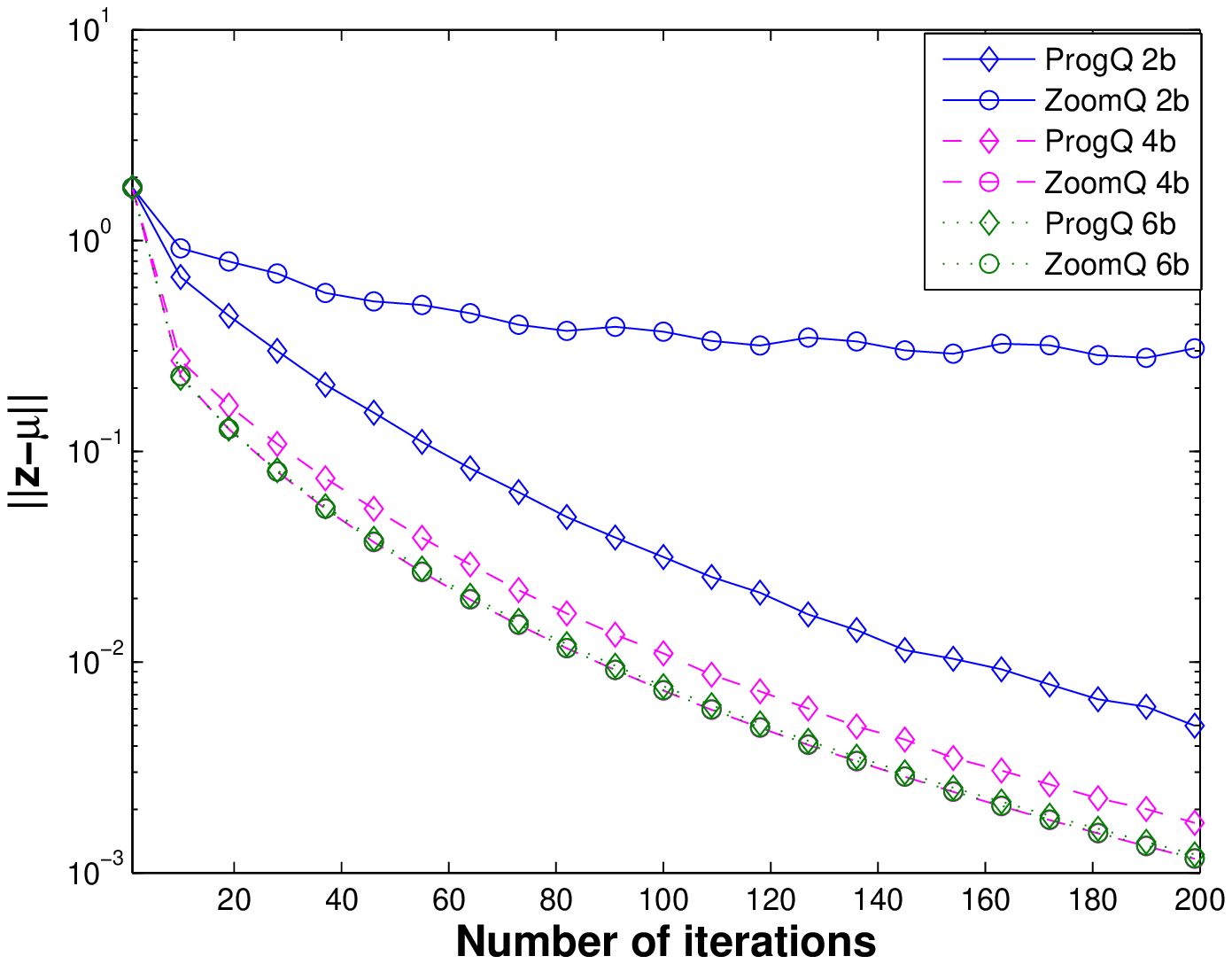}}
\subfigure[Laplacian weights]{\includegraphics[width=0.45\textwidth]{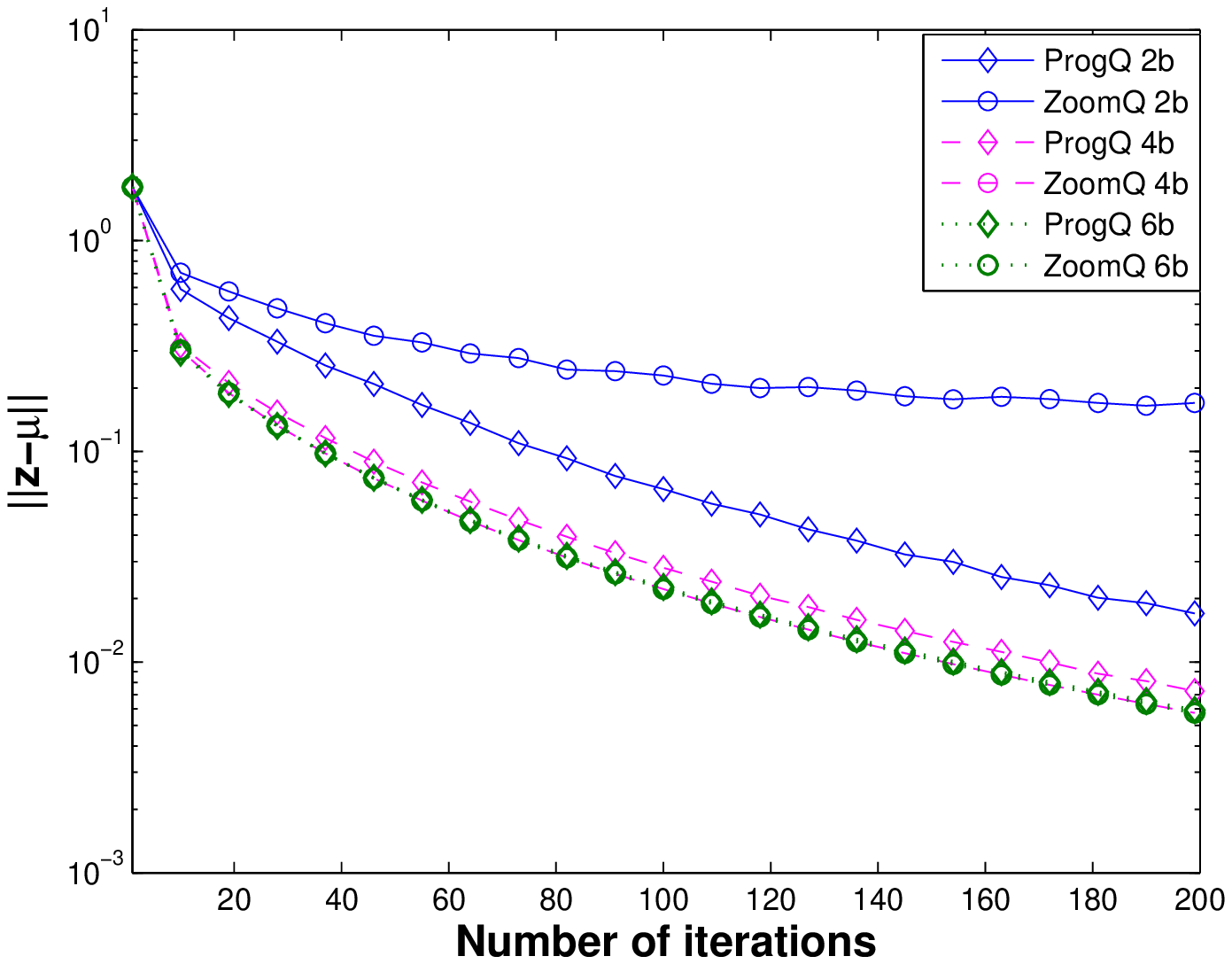}}
\end{center}
\vspace{-0.2cm}
\caption{Average consensus performance of the proposed quantization scheme (ProgQ) vs the zoom-in, zoom-out uniform quantizer (ZoomQ) \cite{carli} for 2, 4 and 6 bits.}
\label{fig:comp3}
\vspace{-0.2cm}
\end{figure*}

\begin{figure*}[tb]
\begin{center}
\subfigure[Metropolis weights]{\includegraphics[width=0.45\textwidth]{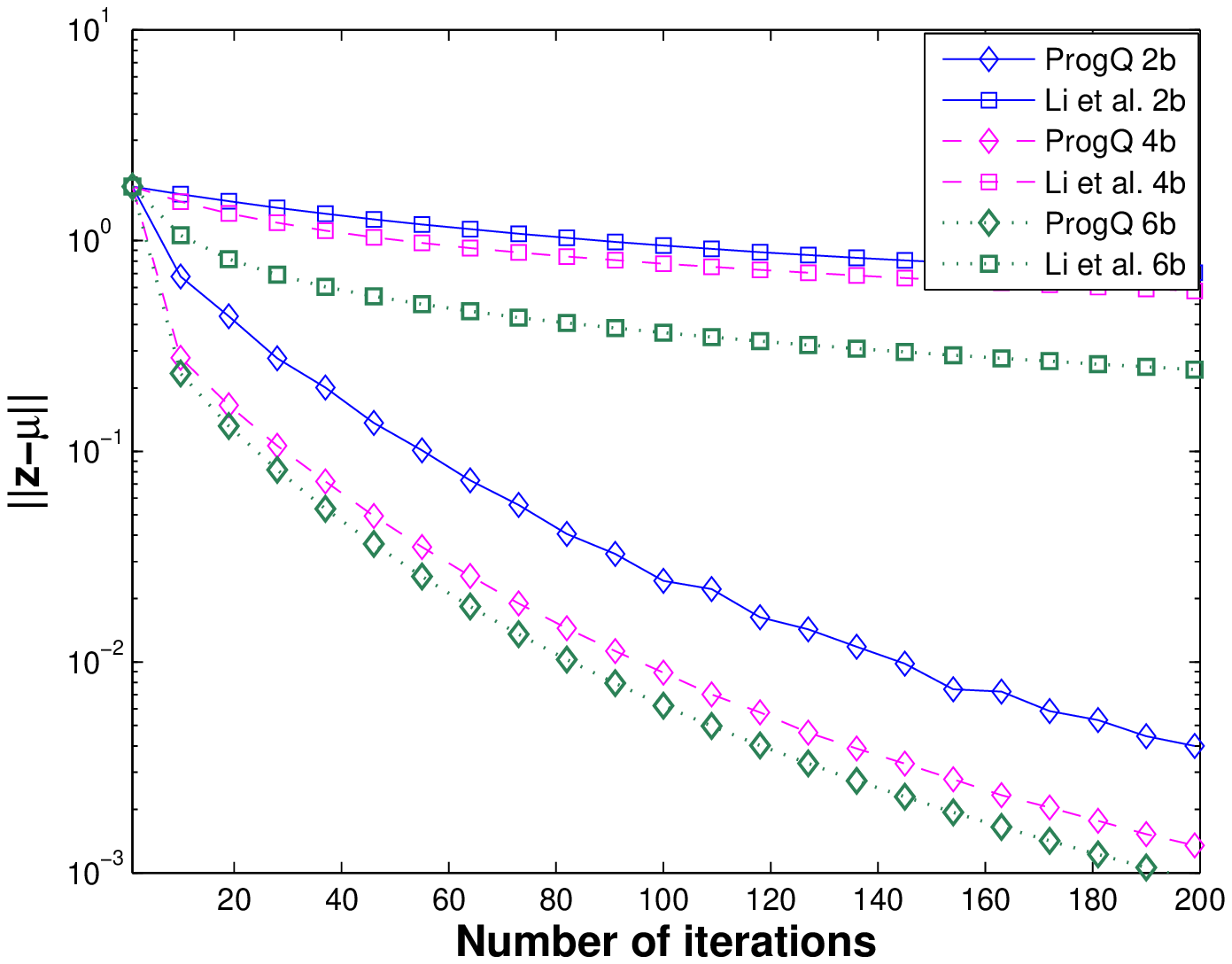}}
\subfigure[Laplacian weights]{\includegraphics[width=0.45\textwidth]{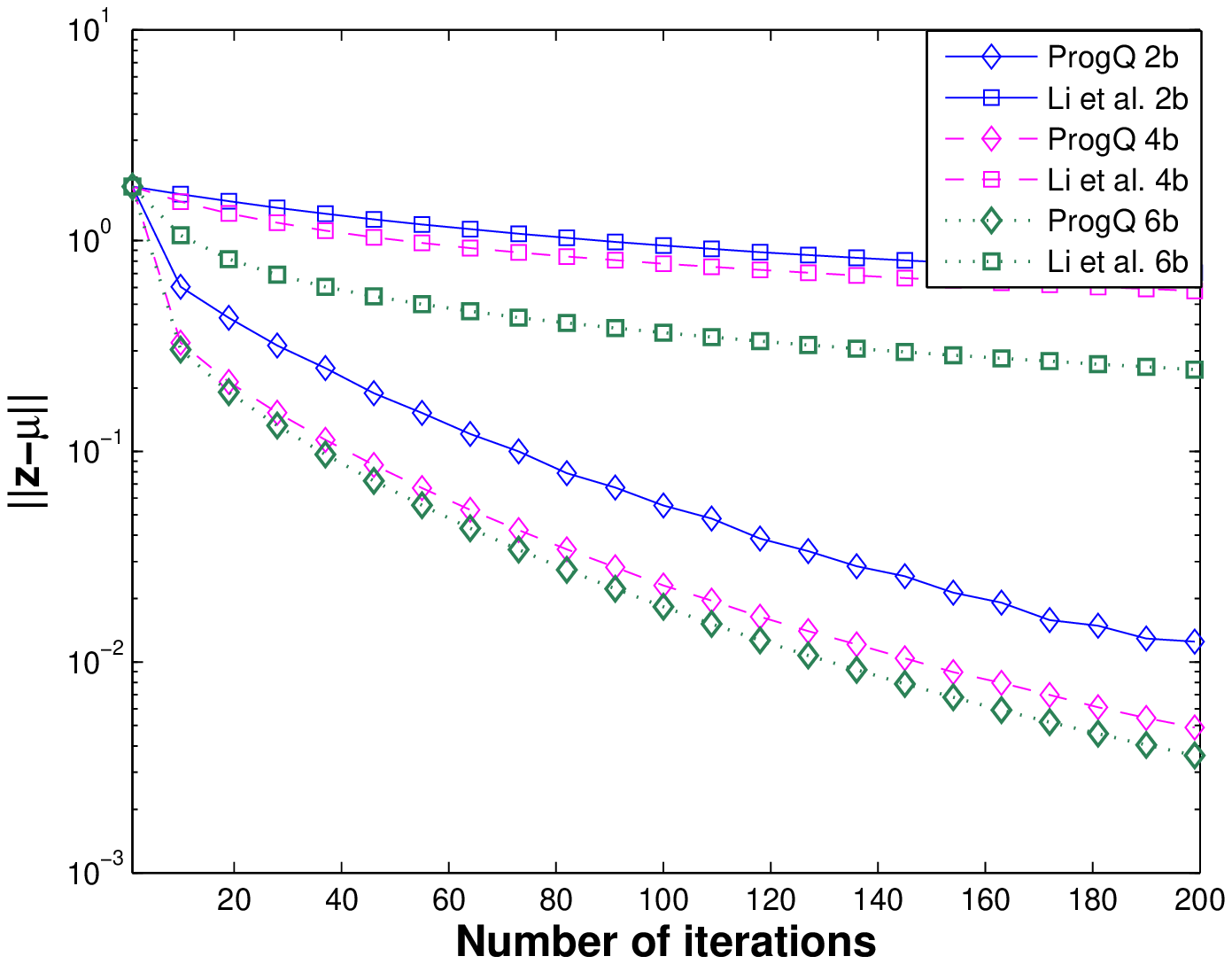}}
\end{center}
\vspace{-0.2cm}
\caption{Average consensus performance of the proposed quantization scheme (ProgQ) vs the quantization scheme proposed in \cite{Li-2011}  (Li et al.) for 2, 4 and 6 bits.}
\label{fig:comp4}
\vspace{-0.2cm}
\end{figure*}

We compare the proposed Progressive Quantizer ('ProgQ') with (a) the adaptive quantizer ('AdaptQ') \cite{Fang09},   (b) the zoom in-zoom out uniform encoder ('ZoomQ')  \cite{carli} and (c) the quantization scheme proposed in \cite{Li-2011} ('Li et al.'). 
In particular, the scheme proposed in \cite{Fang09} is based on the Delta modulation with variable step-size. The step-size is adapted throughout the iterations based on the previously sent bits and a constant $K$. 
However, the scheme is quite sensitive to the value of  $K$ and the performance can deteriorate for non-carefully chosen values.   In our experiments we  choose  $K=1.2$ as defined in \cite{Fang09}. On the other hand,  the differential encoding scheme proposed in \cite{carli} uses a uniform quantizer and the transmitted value is the quantized difference of the current value from the previous estimate, scaled by a factor $f$ that is adapted over time. This factor is similar to the step-size of \cite{Fang09} and it grows or decreases depending on the difference between the new state $z_{t+1}$ and the previously quantized state $\hat{z}_t$. 
The decrease or the increase depends on the constants $k_{in}$ and $k_{out}$ respectively and the way that these constants have to be determined seems to be an open question.  In our experiments we choose the parameters $k_{in}=0.5, k_{out}=2$ and the scaling factor $f_0=0.5$ as defined in \cite{carli}.
Finally, the scheme proposed in \cite{Li-2011}  is also designed  by adapting the scaling function of a difference encoder similar to our quantizer. The value that is quantized at each time step  is the
difference between the new state $z_{t+1}$ and the previously quantized state $\hat{z}_t$ while the scaling function is assumed to have an exponential decay over time. The parameters of the scaling function are defined in an interval form in such a way that the quantizer never saturates. However, one limitation of the scheme in \cite{Li-2011} is  that the algorithm is designed only for the Laplacian weights of the form $W= I-a L$, where the parameter $a$ depends on the number of quantization bits. In order to directly compare the performance of the Progressive Quantizer with the quantization scheme of \cite{Li-2011}, we implement the version of the latter scheme that corresponds to a fixed number of quantization levels (Algorithm 1 in \cite{Li-2011} ). The parameters of the scaling function are representative values that belong to the proposed intervals. The edge weights of the matrix $W$   are computed by  Eqs. (\ref{metropolis}) and  (\ref{laplacian}) for the Progressive Quantizer, the adaptive quantizer and the zoom-in zoom-out quantizer  while  for the quantization scheme of  \cite{Li-2011} they are computed as defined in the corresponding paper, as they depend on the selected parameters of the scaling function. 


We use the same experimental setup as in the previous experiments. Figs \ref{fig:comp2}, \ref{fig:comp3}, \ref{fig:comp4} illustrate the simulation results and show performance comparisons for the quantizers with different bit rates. Notice first that our scheme outperforms the three above mentioned schemes in all the cases.  AdaptQ appears  to saturate especially for a small number of bits.  The performance of ZoomQ seems to be quite good for  4 and 6 bits, but it suffers significantly at low bit rate. On the other hand, the performance of the last scheme (Li et al.) is quite poor even for 6 bits.  This result is quite expected since the proposed intervals for the parameters of the scaling function are too conservative; they are computed  such  that no clipping appears during the iterative consensus algorithm. Moreover, we observe that both the selection of the weight matrix as indicated in \cite{Li-2011} and its dependence on the bit rate penalize even more the convergence rate and the overall performance of the consensus algorithm. 

Our scheme bear some resemblance with these three schemes in the sense that we also propose to adapt a scaling function.  The scaling function has a very specific definition in our case where it represents  the sensors' dynamic range.   Moreover, we impose a consistent decay of the quantizer range size  which is intuitively supported by the increasing correlation of the sensors' states throughout the iterations.  The parameters $\alpha$ and $\gamma$ that determine the  quantization range in our Progressive Quantizer have been carefully designed by taking into consideration both the available number of bits and the topology of the network and they are automatically determined in closed-form from Eqs. (\ref{alpha}) and (\ref{gamma}). Finally, the performance comparison with the scheme proposed in \cite{Li-2011}  confirms  our initial intuition that  very conservative bounds do not necessarily improve the average consensus performance, and that average case analysis is more efficient in practical settings.  

\subsection{  Convergence of the consensus algorithm}
\begin{figure*}[tb]
\begin{center}
\subfigure[]{\label{fig:quantrange} \includegraphics[width=0.45\textwidth]{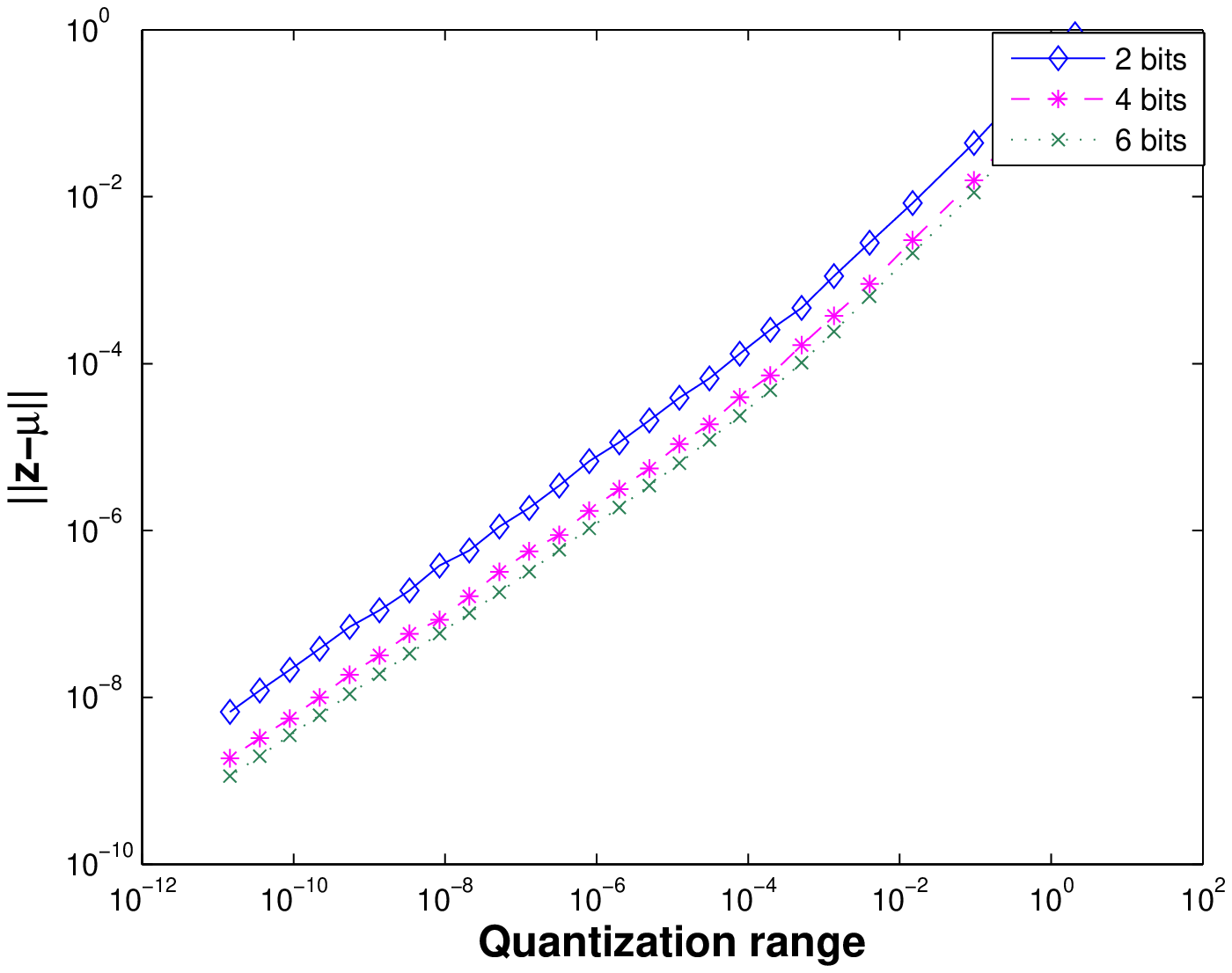}}
\subfigure[]{\label{fig:noisevar} \includegraphics[width=0.45\textwidth]{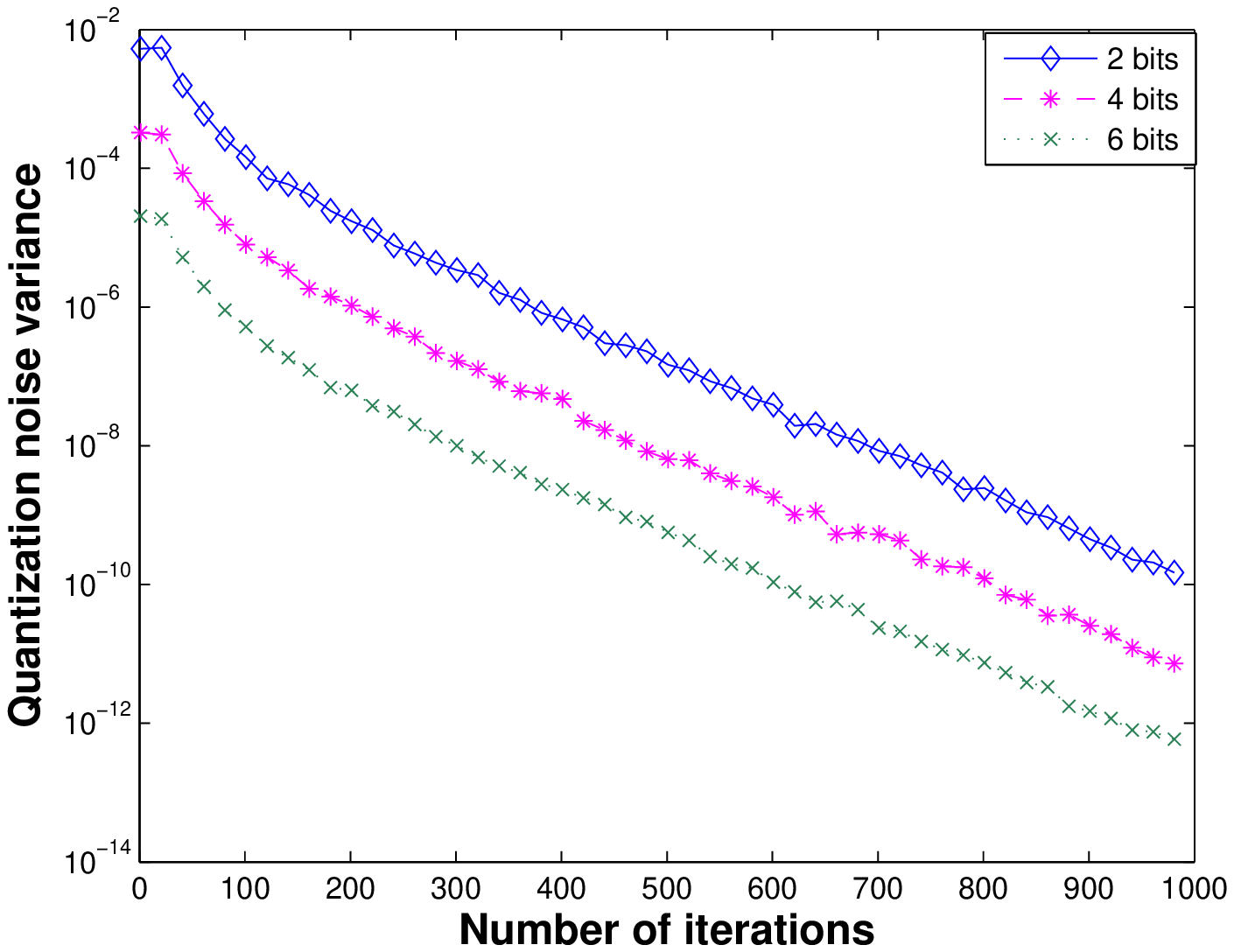}}
\end{center}
\vspace{-0.2cm}
\caption{(a) Evolution of the average consensus performance for an exponentially reducing quantization range. (b) Evolution of the variance of the quantization noise over the iterations for 2, 4 and 6 bits. }
\label{fig:quantrange2}
\vspace{-0.2cm}
\end{figure*}

The assumption that the quantizer saturates complicates significantly the convergence analysis of the proposed algorithm. Our Progressive Quantizer may generate some clipping of the values computed by sensors. We have shown through extensive experimental results that this clipping does not significantly penalize the convergence of the consensus algorithm.   It can even help in case of strong outliers. Moreover, we have observed that the number of clipping, if any, is small and decays to zero as the iteration of the consensus algorithm increases, as long as the parameter $\gamma$ is computed according to Eq.(\ref{gamma}). However, clipping results into some important non-linear effects that are difficult to analyze. In this subsection, we give some intuition about the convergence properties as well as the convergence speed of the Progressive Quantizer.  The simulation results provided in the previous subsections verify that the proposed scheme leads the sensors to converge to the average of their initial values.  We notice first that  the quantization range reduces to zero as time elapses,  leading to an accurate average consensus that is reached independently of the number of bits.  The latter is verified experimentally in Fig.\ref{fig:quantrange}, where we observe that, as the range reduces to zero, the absolute error from the accurate consensus value $\mu$ becomes smaller.  

Moreover, the design of the Progressive Quantizer promotes the decrease of the quantization noise variance over time. By properly decreasing the quantization range, we reduce  the  quantization noise, as long as the computed values to be quantized fall into that range. 
 We finally relate the convergence of the algorithm to the quantization noise in  the following proposition.  Similar results have been shown in \cite{Yildiz08}. 

\begin{proposition}\label{Prop2} Let $\sigma^2_t$ be the sample variance of the quantization noise vector $\epsilon_t$ at iteration $t$.  If  $\sigma_t^2\to0$ as $t\to\infty$,  the sensor nodes converge asymptotically to the true consensus value $\mu$. 
\end{proposition} 
The proof of Proposition \ref{Prop2} is given in Appendix \ref{app4}. 

 Fig.\ref{fig:noisevar} 
verifies that our scheme leads to a quantization noise variance that converges  to zero as time elapses with exactly the same  behavior as the quantization range.  These results are consistent with the ones shown in Fig. \ref{fig:comp1}. It confirms that the  average consensus performance is directly related to the decay of the quantization noise variance and that an accurate consensus is achieved for a variance that converges to zero. Finally, the decay of the noise variance and thus the convergence speed depend on the number of the quantization bits;  more precisely the algorithm converges faster for a large number of bits, which is expected.

\section{Conclusions}
In this paper, we have proposed a novel quantization scheme for solving the average consensus problem when sensors exchange quantized state information. In particular, our scheme is based on progressive reduction of  the range of a uniform quantizer. It leads to progressive refinement of the information exchanged by the sensors while the average consensus algorithm converges.   Simulation results show the effectiveness of our scheme that outperforms other quantized consensus algorithms in terms of convergence rate and accuracy of the computed average. Finally, our quantizer represents a constructive simple solution with effective performance in realistic settings. 

 \appendices 

\section{\textbf{Proof of Proposition \ref{Prop1}}}
\label{app1}

Eq. (\ref{eq11}) implies that
\begin{equation}
{\small
\begin{split}
z_{t+1}-\hat{z}_{t} &= W^{t}(W-I)z_0+(W-2I)\epsilon_t +(W-I)\sum_{s=0}^{t-1}W^s(W-I)\epsilon_{t-s-1}. \label{error2}
\end{split}
  }
\end{equation}
Let us define  $$A_1 := W^{t}(W-I)z_0$$  and $$A_2 := (W-I) \displaystyle \sum_{s=0}^{t-1}W^s(W-I)\epsilon_{t-s-1}+(W-2I)\epsilon_t .$$ Then, we observe that
\begin{equation}
{\small
		\begin{split}
		E[\|z_{t+1}-\hat{z}_{t} \|^2]&=
		   			 E[\|A_1\|^2]+E[A_1^\top A_2]+E[A_2^\top A_1]+E[\|A_2\|^2].
  \end{split}
  \label{error22}
  }
\end{equation}
In the formula above, $\|A_1\|^2$ is a deterministic quantity depending on the (fixed) network topology and the initial states of the sensors. 
We assume that the quantization noise samples $\epsilon_t(i)$ in (\ref{eq:quantZ}) are (spatially and temporally) independent random variables that are uniformly distributed with zero mean and variance $\Delta_t^2/12$, where $\Delta_t$ is the quantization step-size at step $t$. The latter  assumption is widely used in the literature \cite{Yildiz08, Fang09, carli2009} for modeling the quantization noise  and is true under the conditions specified in  \cite{snyder}.  
Due to this independence assumption, the two cross terms  $E[A_1^\top A_2]$ and $E[A_2^\top A_1]$ become zero, and (\ref{error22}) simplifies to
 \begin{equation}
 {\small
		E[\|z_{t+1}-\hat{z}_{t} \|^2]=\|A_1\|^2+E[\|A_2\|^2].
  \label{error3}
  }
\end{equation}
In the sequel, we work out each individual term separately.
First, note that
\begin{equation}
{\small
\begin{split}
\|A_1\|^2&=\|W^t(W-I)z_0\|^2\\
&\overset{(a)}=\|W^t(W-I)z_0-\frac{{\mathbf 1\mathbf{1}^T}}{m}(W-I)z_0\|^2\\
&\overset{(b)}\le \left|\left|W-\frac{{\mathbf1\mathbf{1}^T}}{m}\right|\right|_2^{2t}\|W-I\|^2\|z_0\|^2,
\end{split}
\label{eq:A1}
}
\end{equation}
where (a) follows from the fact that $W$ is doubly stochastic and (b) holds because $\left|\left|W^t-\frac{{\mathbf1\mathbf{1}^T}}{m}\right|\right|_2=\lambda_2^t$, where $\lambda_2:=\rho(W-\frac{{\mathbf1\mathbf{1}^T}}{m})$. Moreover, 
\begin{equation}
{\small
\begin{split}
E[ \|A_2\|^2 ]&= E[ \|(W-I)\sum_{s=0}^{t-1}W^s(W-I)\epsilon_{t-s-1}+(W-2I)\epsilon_t\|^2 ] \\
&\le\|W-I\|^2\sum_{s=0}^{t-1}\|W^s(W-I)\|^2 E[ \|\epsilon_{t-s-1}\|^2] + \|W-2I\|^2 E[ \|\epsilon_{t}\|^2 ],\\
\end{split}
\label{a3}
\notag
}
\end{equation}
where we have again exploited the fact that the quantization noise variables are independent and zero mean.
We compute the expectation of the quantization error norms by exploiting the consistently reduced range of the Progressive Quantizer. In particular, the expectation of the error norm at each time step $t$ can be expressed as $E[\|\epsilon_{t}\|^2]\le mE[\epsilon_{t}(i)^2]\le m\Delta_t^2/12$ \cite{snyder}, where $\Delta_t$ is the quantization step-size at step $t$. Hence, the expectation of the second term is bounded as follows
\begin{equation}
{\small
\begin{split}
E[\|A_2\|^2]
&\le\|W-I\|^2\sum_{s=0}^{t-1}\|W^s(W-I)\|^2m\frac{\Delta_{t-s-1}^2}{12}+\|W-2I\|^2m\frac{\Delta_{t}^2}{12}.\\
\end{split}
\label{a2}
}
\end{equation}
We finally derive an upper-bound for  $E[\|z_{t+1}-\hat{z}_{t} \|^2]$ by combining both (\ref{eq:A1}) and (\ref{a2}) and taking into consideration that $\Delta_t^2 = \left(\frac{S_t}{2^n}\right)^2 $ (see also Eq. (\ref{eq:Delta_t})).  Altogether, we obtain
\begin{equation}
{\small
\begin{split}
E[\|z_{t+1}-\hat{z}_{t} \|^2]&\le \|z_0\|^2 \lambda_2^{2t}\|W-I\|^2+\|W-I\|^2\sum_{s=0}^{t-1}\|W^s(W-I)\|^2m\frac{S_{t-s-1}^2}{2^{2n}\cdot 12}\\ &+ \|W-2I\|^2m\frac{S_{t}^2}{2^{2n}\cdot 12}.
\end{split}\notag
}
\end{equation}
Since the eigenvalues of the matrix $W$ lie in the interval $[-1,1]$, the Proposition \ref{Prop1} follows from the fact that $\|W-I\|=1-\lambda_{\min}$ and $\|W-2I\|=2-\lambda_{\min}$, where $\lambda_{\min}$ is the smallest algebraically eigenvalue of $W$. 
\qed

\section{\textbf{Proof of Proposition \ref{Prop3}}}
\label{app2}

 Firstly, we define a new sequence $P(t)$ to be the upper-bound of $e^{-2\beta_{t}}$.

\begin{definition}\label{def:pt}
The sequence $P(t)$ is defined as:
{\small
\begin{align*}
P(0) &= \frac{(z^{(\max)}_0-z^{(\min)}_0)^{2}}{4}\\
P(1) &= \|z_0\|_\infty^{2}(1-\lambda_{\min})^{2}+\frac{(2-\lambda_{\min})^{2}}{2^{2n}\cdot3}P(0)\\
P(t) &= \|z_0\|_\infty^{2}(1-\lambda_{\min})^{2}\lambda^{2(t-1)}_2+(1-\lambda_{\min})^4\sum^{t-2}_{s=0}\lambda^{2s}_2\frac{P(t-2-s)}{2^{2n}\cdot3}+(2-\lambda_{\min})^{2}\frac{P(t-1)}{2^{2n}\cdot3}, \quad t \geq 2.
\end{align*}
}
Moreover, 
{\small
\begin{align*}
P(0) &= e^{-2\beta_{0}}\\
P(1) &= e^{-2\beta_{1}}\\
P(t) &\geq e^{-2\beta_{t}}, \quad t \geq 2.
\end{align*}
}
\end{definition}


%
%
%

We can then write the sequence in the following form
{\small
\begin{align}
P(t+1)=(c+\gamma)P(t)+(b-c\gamma)P(t-1), 
\end{align}
}
where $c=\frac{(2-\lambda_{\min})^{2}}{2^{2n}\cdot3}$,   $b=\frac{(1-\lambda_{\min})^4}{2^{2n}\cdot3}$ and $\gamma=\lambda_2^2$ are positive constants. Expressing the above sequence in a matrix form we obtain

{\small
\begin{align}\label{eq:13}
\begin{bmatrix}	
        P(t+1) \\
        P(t) 
\end{bmatrix}
=\underbrace{\begin{bmatrix}	
        c+\gamma & b-c\gamma \\
        1        & 0 
\end{bmatrix}}_{A}\begin{bmatrix}	
        P(t) \\
        P(t-1) 
\end{bmatrix}.
\end{align}
}

The eigenvalues of the matrix $A$ are $\frac{c+\gamma\pm \sqrt{(c-\gamma)^{2}+4b}}{2}$ and  
the dynamical system (\ref{eq:13}) converges to zero if its eigenvalues are strictly smaller than 1 \cite{luenberger}. Since $c$, $b$ and $\gamma$ are three positive constants, it is enough to find conditions that guarantee that $\frac{c+\gamma+ \sqrt{(c-\gamma)^{2}+4b}}{2}<1$. After substitution of the values of the constants $c$, $b$ and $\gamma$, it leads to the following inequality 


%
%


{\small
\begin{align*}
\frac{(2-\lambda_{\min})^2}{3\cdot 2^{2n}} + \lambda^{2}_2 + \sqrt{(\frac{(2-\lambda_{\min})^2}{3\cdot 2^{2n}} - \lambda^{2}_2)^{2}+4\frac{(1-\lambda_{\min})^4}{3\cdot 2^{2n}}}< 2.
\end{align*}
}


Equivalently, we can write 
{\small
\begin{align}\label{eq:14}
\frac{4(1-\lambda_{\min})^4-4(\lambda^{2}_2-1)(2-\lambda_{\min})^{2}}{3\cdot 2^{2n}} < (2\lambda^{2}_2-2)^{2}-2\lambda^{2}_2(2\lambda^{2}_2-2),
\end{align}
}
which indicates that $P(t)$ converges to zero if the following condition 
{\small
\begin{align}\label{eq:15}
\frac{(1-\lambda_{\min})^4}{(1-\lambda^{2}_2)} +(2-\lambda_{\min})^{2} < 3\cdot 2^{2n}
\end{align}
}
is satisfied. Since the sequence $P(t)$ is an upper-bound of the positive sequence $e^{-2\beta_{t}}$,  if the condition (\ref{eq:15}) is satisfied,  $e^{-2\beta_{t}}$ also converges to zero. 
\qed



\section{\textbf{Proof of Proposition \ref{Prop4}}}
\label{app3}
Recall from (\ref{eq:A1}) that $\|W^{t}(W-I)\|^2\le\|W-\frac{{\mathbf{1}\mathbf{1}^T}}{m}\|^{2t}\|W-I\|^2=\lambda_2^{2t}(1-\lambda_{\min})^2$. Moreover, under the condition specified in Proposition \ref{Prop3}, we have that $\lim_{t\rightarrow \infty} e^{-\beta_{t}}=0$.  Given $\epsilon>0$, there exists a time instant $k>0$ such that $\forall t\ge k$, $e^{-\beta_{t}}<\epsilon$. Then,  
\begin{equation}
\small{
\begin{split}
\sum_{s=0}^{t-1}\|W^s(W-I)\|^2{e^{-2\beta_{t-s-1}}}&\le\sum_{s=0}^{t-1}\|W-\frac{{\mathbf{1}\mathbf{1}^T}}{m}\|^{2s}\|W-I\|^2e^{-2\beta_{t-s-1}}\\
&=\|W-I\|^2\sum_{s=0}^{t-1}\|W-\frac{{\mathbf{1}\mathbf{1}^T}}{m}\|^{2s}e^{-2\beta_{t-s-1}}.
\end{split}
}
\end{equation}
Moreover, since the eigenvalues of the matrix $W$ lie in the interval $[-1,1]$ we have that $\|W-I\|=1-\lambda_{\min}\le2$. Hence, using these observations and setting $l=t-s-1$, the above inequality can be written as
\begin{equation}
\small{
\begin{split}
&\sum_{s=0}^{t-1}\|W^s(W-I)\|^2{e^{-2\beta_{t-s-1}}}\le4 \sum_{l=0}^{t-1}\|W-\frac{{\mathbf{1}\mathbf{1}^T}}{m}\|^{2(t-l-1)}e^{-2\beta_{l}}\\
&<4\sum_{l=0}^{k}\|W-\frac{{\mathbf{1}\mathbf{1}^T}}{m}\|^{2(t-l-1)}e^{-2\beta_{l}}+4\sum_{l=k+1}^{t-1}\|W-\frac{{\mathbf{1}\mathbf{1}^T}}{m}\|^{2(t-l-1)}\epsilon^2.\\
\end{split}
}
\end{equation}
 We notice that, since $\|W-\frac{\boldsymbol{1}\boldsymbol{1}^T}{m}\|<1$, $e^{-\beta_t}\le \delta, \forall t$ and $k$ is finite, the first term of the right-hand side of the inequality converges to 0 for $t\to\infty$.  Finally, we can write 
 \begin{equation}
\small{
\begin{split}
\lim_{t\rightarrow \infty} \sum_{s=0}^{t-1}\|W^s(W-I)\|^2{e^{-2\beta_{t-s-1}}}&< \lim_{t\rightarrow \infty} 4\sum_{l=k+1}^{t-1}\|W-\frac{{\mathbf{1}\mathbf{1}^T}}{m}\|^{2(t-l-1)}\epsilon^2\\
&<\lim_{t\rightarrow \infty} 4C\epsilon^2\to 0, \mbox{ for } \epsilon\to 0,
\end{split}
}
\end{equation}
 where we have used the fact that $\lim_{t\to\infty}\sum_{l=k+1}^{t-1}\|W-\frac{\boldsymbol{1}\boldsymbol{1}^T}{m}\|^{2(t-l-1)}$ converges to a constant $C<\infty$.

\qed

\section{\textbf{Proof of Proposition \ref{Prop2}}}
\label{app4}
The proof of Proposition \ref{Prop2} is similar to the one of Proposition \ref{Prop4}. Assume that the sample variance $\sigma_t^2$ of the quantization noise components converges to zero after some specific iterations. In particular, given $\delta>0$, there exists a time instant $k>0$ such that $\forall t\ge k$, $\sigma^2_t<\delta$. Using Eq.(\ref{eq11}), the Euclidean deviation of the node states can be expressed as 
\begin{equation}
{\small
\begin{split}
\|z_{t+1}-\frac{\boldsymbol{1}\boldsymbol{1}^T}{m}z_0 \|&=\|W^{t+1}z_0+\sum_{s=0}^{t}W^s(W-I)\epsilon_{t-s}-\frac{\boldsymbol{1}\boldsymbol{1}^T}{m}z_0 \|\\
&\le \|W^{t+1}z_0-\frac{\boldsymbol{1}\boldsymbol{1}^T}{m}z_0\|+\sum_{s=0}^{t}\|W^s(W-I)\epsilon_{t-s}\|\\
&\le\|W-\frac{\boldsymbol{1}\boldsymbol{1}^T}{m}\|^{t+1}\|z_0-\frac{\boldsymbol{1}\boldsymbol{1}^T}{m}z_0\| \\& + \sum_{s=0}^{t}\|(W-\frac{\boldsymbol{1}\boldsymbol{1}^T}{m})^s(W-I)(\epsilon_{t-s}-\frac{\boldsymbol{1}\boldsymbol{1}^T}{m}\epsilon_{t-s})\|\\
&\le\|W-\frac{\boldsymbol{1}\boldsymbol{1}^T}{m}\|^{t+1}\|z_0-\frac{\boldsymbol{1}\boldsymbol{1}^T}{m}z_0\| +\\ & +\sum_{s=0}^{t}\|W-\frac{\boldsymbol{1}\boldsymbol{1}^T}{m}\|^s\|W-I\|\|\epsilon_{t-s}-\frac{\boldsymbol{1}\boldsymbol{1}^T}{m}\epsilon_{t-s}\|,
\end{split}
\label{a4}
\notag
}
\end{equation}
where we have used again the properties of the matrix $W$ as defined in Eq. (\ref{eq:condW}). 
Notice that $\|\epsilon_{t-s}-\frac{\boldsymbol{1}\boldsymbol{1}^T}{m}\epsilon_{t-s}\|$ can be written in terms of the sample variance $\sigma^2_{t-s}$ of the quantization noise  at iteration $t-s$ such as $\|\epsilon_{t-s}-\frac{\boldsymbol{1}\boldsymbol{1}^T}{m}\epsilon_{t-s}\|=\sqrt{m\sigma^2_{t-s}}$. Moreover, 
using the fact that $\|W-I\|=1-\lambda_{\min}\le 2$ and setting $l=t-s$, the above inequality can be written as
\begin{equation}
{\small
\begin{split}
\|z_{t+1}-\frac{\boldsymbol{1}\boldsymbol{1}^T}{m}z_0 \|&\le \|W-\frac{\boldsymbol{1}\boldsymbol{1}^T}{m}\|^{t+1}\|z_0-\frac{\boldsymbol{1}\boldsymbol{1}^T}{m}z_0\|+2\sqrt{m}\sum_{l=0}^{t}\|W-\frac{\boldsymbol{1}\boldsymbol{1}^T}{m}\|^{t-l}\sqrt{\sigma_l^2}\\
&=\|W-\frac{\boldsymbol{1}\boldsymbol{1}^T}{m}\|^{t+1}\|z_0-\frac{\boldsymbol{1}\boldsymbol{1}^T}{m}z_0\|+2\sqrt{m}(\sum_{l=0}^{k}\|W-\frac{\boldsymbol{1}\boldsymbol{1}^T}{m}\|^{t-l}\sqrt{\sigma_l^2}\\
&+\sum_{l=k+1}^{t}\|W-\frac{\boldsymbol{1}\boldsymbol{1}^T}{m}\|^{t-l}\sqrt{\sigma_l^2}).
\end{split}
\label{a5}
\notag
}
\end{equation}
We notice that since $\|W-\frac{\boldsymbol{1}\boldsymbol{1}^T}{m}\|<1$ and $k$ is finite, the first two terms of the right-hand side of the inequality converge to 0 for $t\to\infty$. Furthermore, by assumption we have that $\sigma^2_t<\delta, \forall t>k$, which implies that  for $t\to\infty$, we have
\begin{equation}
{\small
\lim_{t\to\infty}\|z_{t+1}-\frac{\boldsymbol{1}\boldsymbol{1}^T}{m}z_0 \|<2\sqrt{m}\delta\lim_{t\to\infty}\sum_{l=k+1}^{t}\|W-\frac{\boldsymbol{1}\boldsymbol{1}^T}{m}\|^{t-l}<2\sqrt{m}\delta C,
\notag
}
\end{equation}
where again we have used the fact that $\lim_{t\to\infty}\sum_{l=k+1}^{t}\|W-\frac{\boldsymbol{1}\boldsymbol{1}^T}{m}\|^{t-l}$ converges to a constant $C<\infty$. 
Finally, as the variance converges to 0, the sensors reach a consensus on the average $\mu=\frac{\boldsymbol{1}\boldsymbol{1}^T}{m}z_0$ i.e., 
\begin{equation}
{\small
\lim_{t\to\infty}\|z_{t+1}-\frac{\boldsymbol{1}\boldsymbol{1}^T}{m}z_0 \|\to0 \mbox{ for } \delta\to 0. 
\notag
}
\end{equation}

\qed














\bibliographystyle{IEEEtran}
\bibliography{mybibfile}

\end{document}